\documentclass[a4paper,fleqn]{cas-sc}
\usepackage[numbers]{natbib}
\usepackage{hyperref}

\newlength{\colwidthdc}
\def\tsc#1{\csdef{#1}{\textsc{\lowercase{#1}}\xspace}}
\tsc{WGM}
\tsc{QE}
\tsc{EP}
\tsc{PMS}
\tsc{BEC}
\tsc{DE}
\begin{document}
\let\printorcid\relax
\let\WriteBookmarks\relax
\def\floatpagepagefraction{1}
\def\textpagefraction{.001}
%\shorttitle{Leveraging social media news}
\shortauthors{Dali Zhang et~al.}

%\title [mode = title]{A glass-based coded mask telescope module for soft X-ray imaging}                      
\title [mode = title]{Lightweight Soft X-ray Imager (LSXI) with glass-based coded mask}                      

\author[1]{Dali Zhang}[ type=editor,
                        auid=000,bioid=1,
                        orcid=0000-0003-4311-5804]
\cormark[1]
\ead{Zhangdl@ihep.ac.cn}

%\address[1]{, Street 129, 1043 NX Amsterdam, The Netherlands}
\affiliation[1]{organization={State Key Laboratory of Particle Astrophysics, Institute of High Energy Physics},
                addressline={ Chinese Academy of Sciences.}, 
 %               city={Trivandrum},
%               citysep={}, % Uncomment if no comma needed between city and postcode
                postcode={100049}, 
                state={Beijing},
                country={China}}

\affiliation[2]{organization={National Astronomical Observatories, Chinese Academy of Sciences},
             %   addressline={Street 29}, 
                postcode={100101}, 
              %  postcodesep={}, 
                city={Beijing},
                country={China}}

\affiliation[3]{organization={University of Chinese Academy of Sciences},
               % addressline={Street 15}, 
                %city={Jabaldesh},
                postcode={100049}, 
                state={Beijing}, 
                country={China}}  

\affiliation[4]{organization={TIANFU Cosmic Ray Research Center},
               % addressline={Street 15}, 
                city={Chengdu},
                postcode={610000}, 
                state={Sichuan}, 
                country={China}} 

\affiliation[5]{organization={School of Materials Science and Engineering,NanChang University},
               % addressline={Street 15}, 
                %city={Jabaldesh},
                postcode={330031}, 
                state={Nanchang}, 
                country={China}}   

\affiliation[6]{organization={North Night Vision Technology (Nanjing) Research Institute Co.},
               % addressline={Street 15}, 
                %city={Jabaldesh},
                postcode={210000}, 
                state={Nanjing}, 
                country={China}} 
                
\affiliation[7]{organization={School of Physics, Nankai University},
               % addressline={Street 15}, 
                %city={Jabaldesh},
                postcode={300350}, 
                state={Tianjin}, 
                country={China}}  

\author[1,4]{Chao Zheng}
\author[3]{Lu Wang}
\author[7]{Haopeng Li}
\author[3]{Jinpeng Zhang}

\author[1]{Xinqiao Li}
\cormark[2]
\ead{lixq@ihep.ac.cn}

\author[1]{Shaolin Xiong}

\author[6]{Longhui Li}
\cormark[1]
\ead{bnullh@163.com}
\author[1]{Zhenghua An}
\author[1]{Sheng Yang}
\author[6]{Xiaoqing Cong}

\author[2,3]{Zhixing Ling}
\author[1]{Hailing Qin}
\author[1]{Xiangyang Wen}
\author[1]{Ke Gong}
\author[1]{Yaqing Liu}
\author[1]{Xiaojing Liu}
\author[1]{Xiang Ma}
\author[1]{Xiaoyun Zhao}
\author[1]{Yanbing Xu}
\author[7]{Junhao Yin}
\author[5]{Dejun Gong}

\author[1,3]{Jiacong Liu}
\author[1,3]{Chenwei Wang}
\author[1]{Min Gao}

\author[1]{Jinzhou Wang}

\cortext[cor1]{Corresponding author}
\cortext[cor2]{Principal corresponding author}

\fntext[fn1]{Equal contribution to this work.}

\begin{abstract}
The coded mask technique has been widely used in X-ray and Gamma-ray imagers, especially in space astronomy. However, the traditional design of coded mask imagers usually has problems with large size and heavy weight. Here, we propose a novel design for a coded mask imager made of glass based on microchannel plate (MCP) technology, making it lightweight \textcolor{black}{(within 1 kg)}, compact, and self-supporting, which is very suitable for space exploration satellites. The design is initially demonstrated by reconstructing encoded patterns from measurements at the X-ray beamline. Monte Carlo simulations of the module integrated into a 6U CubeSat are conducted to assess its in-orbit detector performance, including sensitivity and source localization.
\end{abstract}


\maketitle

\section{Introduction}

\par X-ray astronomy is a branch of observational astrophysics that investigates celestial sources via their X-ray emission \cite{Wilkes2022}. Because Earth’s atmosphere is essentially opaque to X-rays, all such observations must be performed from satellite platforms. Since the launches of Chandra \cite{Weisskopf2000} and XMM-Newton \cite{Jansen2001}, X-ray studies have unveiled a wealth of astrophysical processes. The Gravitational Wave High-energy Electromagnetic Counterpart All-sky Monitor (GECAM)\cite{GECAM,GECAMGRD} is an all-sky mission designed to search for electromagnetic counterparts to gravitational wave events. Its primary scientific objective is to observe and investigate a range of high-energy transient phenomena, such as Gamma-ray bursts, soft Gamma-ray repeaters, X-ray binaries, and related sources. The GECAM mission has expanded into a constellation of four instruments (GECAM-A/B, GECAM-C \cite{Zheng2024Ground,GECAM_C_ZHANG2023168586}, and GECAM-D). Among them, GECAM-D, also known as the Gamma-ray Transient Monitor (GTM), is the first Gamma-ray detector to operate in a distant retrograde orbit (DRO)  \cite{GTMLiu2026,IEEEZhang2026}. The Gamma-ray probe (GTP) on GTM \cite{feng_2024_GECAMDGTP} is capable of precise measurements of Gamma-ray spectra and temporal evolution. However, the GTP itself has no intrinsic imaging capability and relies on a localization algorithm that depends on the count rates from all GTP units. Due to payload mass limitations, only five GTPs can be installed, resulting in relatively coarse localization accuracy \textcolor{black}{ranges from several degrees to tens of degrees}. To achieve better localization, a compact and lightweight soft X-ray imager (LSXI) needs to be further developed.

\par Contemporary X-ray telescope payloads place a growing emphasis on lightweight designs for small platforms, enabling stricter mass and volume constraints and accelerated instrument development cycles. Wide-field monitoring missions like the Einstein Probe (EP) \cite{Yuan:2015y8} exemplify a trend toward lightweight instruments that use innovative lobster-eye micro-pore (MPO) technology to conduct a wide-field, high-sensitivity survey of the dynamic X-ray sky. Collimated telescopes of the EXTP type \cite{LIMCP}, for instance, achieve mechanical simplicity through modular optical paths using microchannel plate (MCP) technology.
\par
\begin{figure}
  \centering
    \includegraphics[width=0.8\colwidthdc]{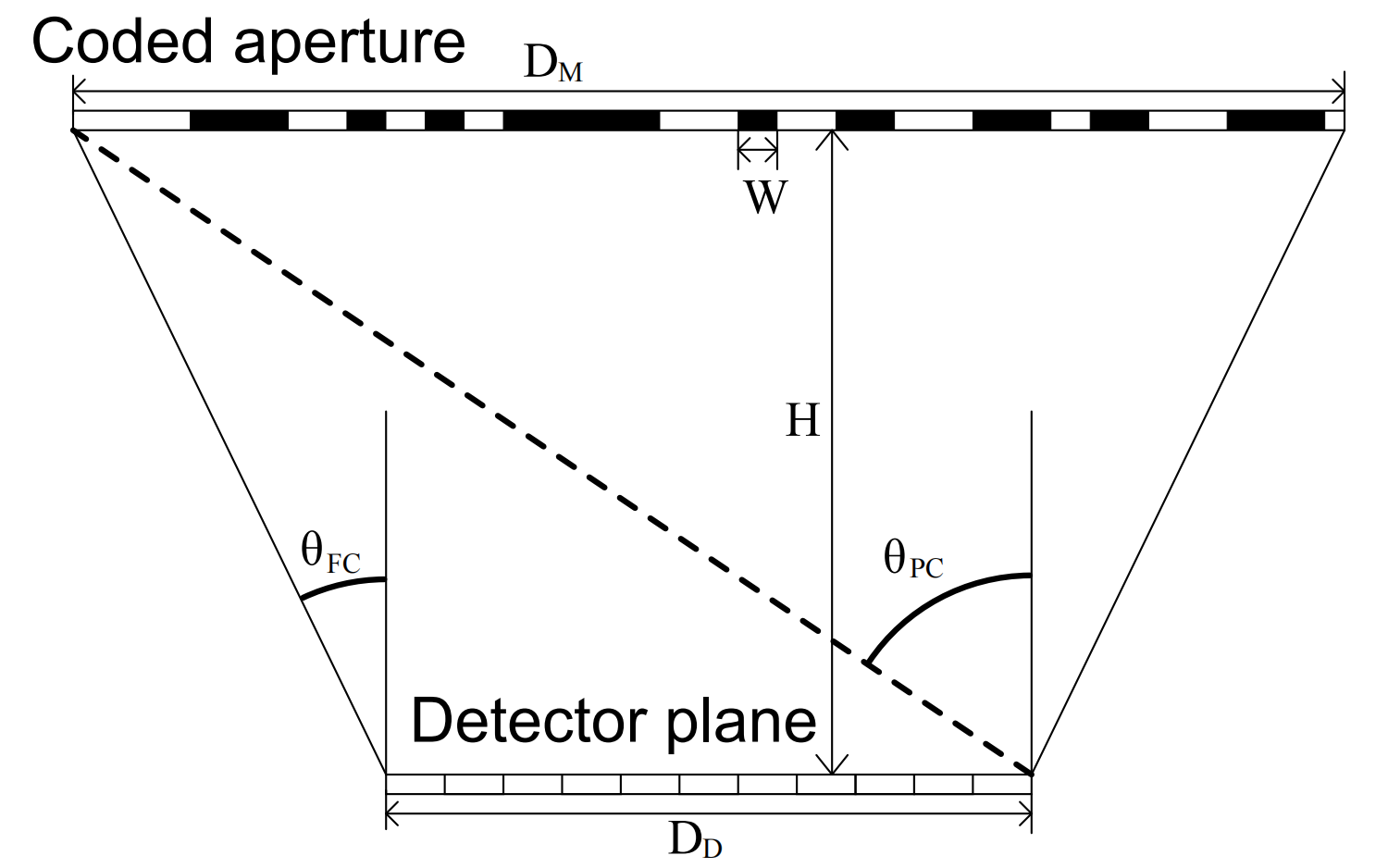}
    \caption{Illustration of a coded-mask telescope. The mask M has a dimension D$_{M}$ that is twice the detector dimension D$_{D}$. H is the distance from the coded mask to the detector plane. W is the mask element size. The fully coded field of view $\theta_{FC}$ (FCFOV) and partially coded field of view $\theta_{PC}$ (PCFOV) are marked.}
    \label{fig:CodeImaging}
\end{figure}

 Coded-mask telescopes are crucial X-ray imagers in X-ray and Gamma-ray astrophysics, especially for probing high-energy phenomena \textcolor{black}{including multiple astrophysical X-ray/Gamma-ray sources}. Representative coded-mask satellite telescopes include INTEGRAL/IBIS \cite{INTEGRAL}, Swift/BAT \cite{Swift}, and SVOM/ECLAIRs \cite{ECLAIRs}. Their basic operating principle is to record the X-ray shadow cast by a coded mask composed of transparent and opaque elements, producing a characteristic shadow pattern on a downstream position-sensitive detector (as shown in Fig. \ref{fig:CodeImaging}). By applying appropriate decoding algorithms, localization can be performed. The coded masks in these telescopes are fabricated from high-density metals (e.g., tungsten), and the total mass of the detector assembly reaches several hundred kilograms. 
\par Lightweight X-ray telescope designs have been achieved for X-ray focusing (MPO of EP) and collimation (MCP of EXTP), but a lightweight design for coded-mask telescopes is not yet available. In a conventional coded-mask telescope, a supporting structure is placed behind the open elements of the mask to enhance mechanical strength, and its attenuation of incident high-energy gamma rays can be ignored. For soft X-rays, however, the open regions of the mask elements must remain completely free of obstruction. This requirement calls for a self-supporting mask architecture.

\begin{figure}
  \centering
    \includegraphics[width=\colwidthdc]{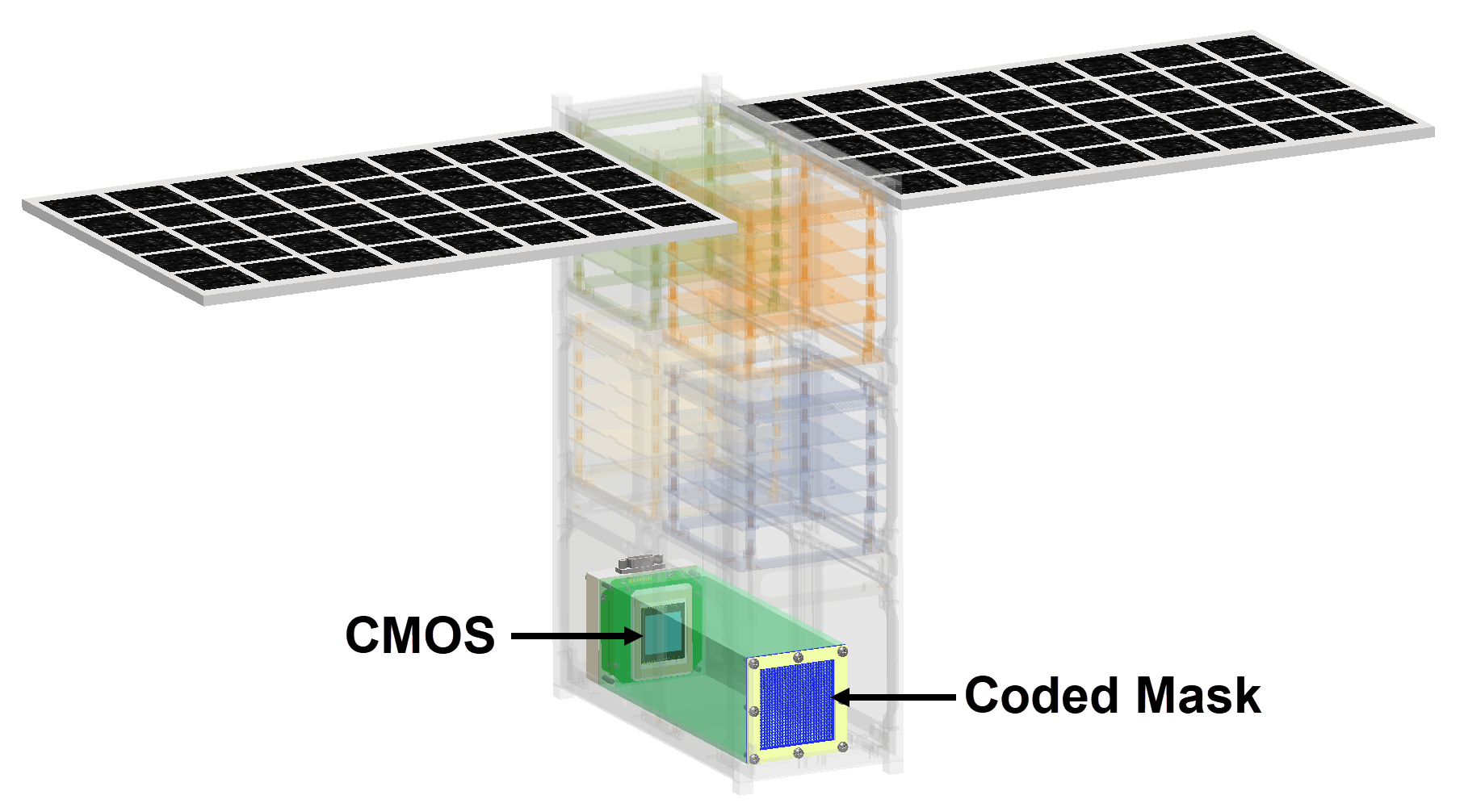}
    \caption{Structural layout of the coded mask telescope module mounted on a 6U CubeSat. \textcolor{black}{The green shadow area is the 1 mm aluminum shielding.}}
    \label{fig:SatelliteStructure}
\end{figure}

\par \textcolor{black}{To our knowledge, this is the first reported glass coded mask telescope prototype based on microchannel plate (MCP) technology, resulting in a lightweight, self-supporting design.} The results presented in this paper consider only the scenario of a 6U CubeSat payload in DRO orbit as an example. \textcolor{black}{The detector is not limited to the satellite payload described in this paper; it can be flexibly adjusted according to the specific payload space available and can be used for X-ray astronomy observation in low Earth orbit or deep space orbit. The rest of this paper is structured as follows: A brief introduction to the telescope design of the 6U CubeSat is given in Section 2. Then, in Section 3, we present the manufacturing process of the coded mask. The beam-test results are reported in Section 4, and performance estimation is reported in Section 5. Finally, we conclude the design and provide an outlook for future work in Section 6.}
%%%%%%%%%%%%%%%%%%%%%%%%%%
\section{Overall Design of LSXI}
\par To depict the idea of LSXI, a single detector module onto a 6U CubeSat is shown in Fig. \ref{fig:SatelliteStructure}. It can be further applied to various nano-satellites because of its compact size and lightweight\textcolor{black}{, while also allowing multi-module combination.} The 6U CubeSat setup of the LSXI module has dimensions of 19 cm in height, 5 cm in length, and 5 cm in width. The coded mask is positioned 16 cm away from the detector plane. This \textcolor{black}{LSXI is mainly composed of a glass-based mask and a CMOS detector}. The CMOS sensor used is the GSENSE-G400 from GPIXEL \cite{G400Wang_2019}, featuring an active area of 22.528 mm × 22.528 mm and a resolution of \textcolor{black}{2048 × 2048} pixels. This CMOS is sensitive to X-rays in the 0.5–8 keV energy range \textcolor{black}{The glass-based coded mask is developed by North Night Vision Technology (Nanjing) Research Institute Co., Ltd. } LSXI provides a fully coded field of view \textcolor{black}{±3.93°} and a partially coded field of view \textcolor{black}{±11.65°}. The total weight of the telescope unit is within 1 kg \textcolor{black}{including it's whole structure}. 
%\section{The design and manufacturing processes of glass-based coded masks}
\section{Glass-based Coded Mask}
\par
\begin{figure}
  \centering
    \includegraphics[width=\colwidthdc]{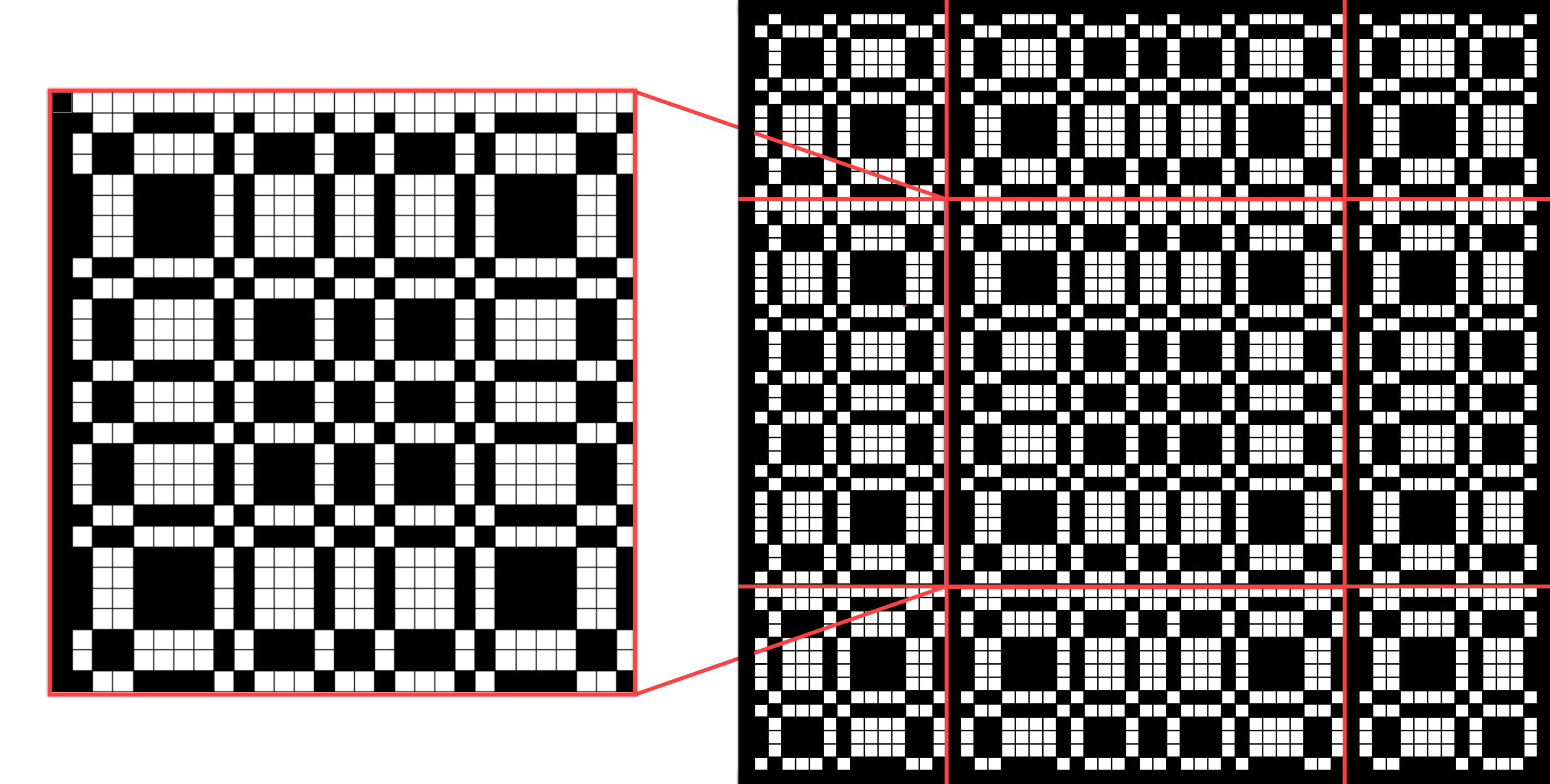}
    \caption{Left: The centralized basic MURA pattern. Right: Its 2$\times$2 arrangement. It has a self-supporting grid structure in the open area.}
    \label{fig:MURAPattern}
\end{figure}

 Coded-mask telescopes typically employ either random mask patterns, as implemented in SWIFT/BAT and SVOM/ECLAIR, or MURA-type patterns, as used in INTEGRAL/IBIS. In contrast to random patterns, MURA patterns avoid introducing inherent mask noise. Therefore, the glass coded mask employed in this study utilizes the MURA pattern, as illustrated in Fig. \ref{fig:MURAPattern}. Due to the micro self-supporting grids of the mask element, the glass mask has an open fraction of 42\%, compared with the ideal open fraction of 50\%. The mask thickness is 0.9 mm, its mask element size is around 0.71 mm, with dimensions of 57 $\times$ 57. The main components are SiO$_{2}$ and Pb, with a density of 3.7 g/$cm^{3}$. 

%\begin{table}
%    \centering
%    \caption{Coded mask characteristics.}
%    \label{tab:Coded mask characteristics.}
%    \renewcommand{\arraystretch}{1.5} % 调整这个值来控制行间距，1.5是1.5倍间距
%    \begin{tabular}{|l|l|}
%        \hline
%        \multicolumn{2}{|c|}{Parameter} \\ \hline
%        Coded mask size & 44 mm$\times$44 mm \\ \hline
%        Open area ratio & ~42\% \\ \hline
%        Thickness & 0.9 mm \\ \hline
%        Main material & SiO$_{2}$+Pb \\ \hline
%        Pb content & 35\% \\ \hline
%        Density & 3.7 g/cm\textsuperscript{3} \\ \hline
%        Mask element size & 0.713 - 0.776 mm \\ \hline
%        Coded mask pattern & MURA \\ \hline
%        Dimension & 57$\times$57 \\ \hline
%    \end{tabular}
%\end{table}
%
%\begin{table}[width=.9\linewidth,cols=4,pos=h]
%\caption{This is a test caption. This is a test caption. This is a test
%caption. This is a test caption. Use \{table*\} instead of \{table\} if you
%want a two column spanned table.}\label{tbl1}
%\begin{tabular*}{\tblwidth}{@{} LLLL@{} }
%\toprule
%Col 1 & Col 2 & Col 3 & Col4\\
%\midrule
%12345 & 12345 & 123 & 12345 \\
%12345 & 12345 & 123 & 12345 \\
%12345 & 12345 & 123 & 12345 \\
%12345 & 12345 & 123 & 12345 \\
%12345 & 12345 & 123 & 12345 \\
%\bottomrule
%\end{tabular*}
%\end{table}

\par
\begin{figure}
  \centering
    \includegraphics[width=\colwidthdc]{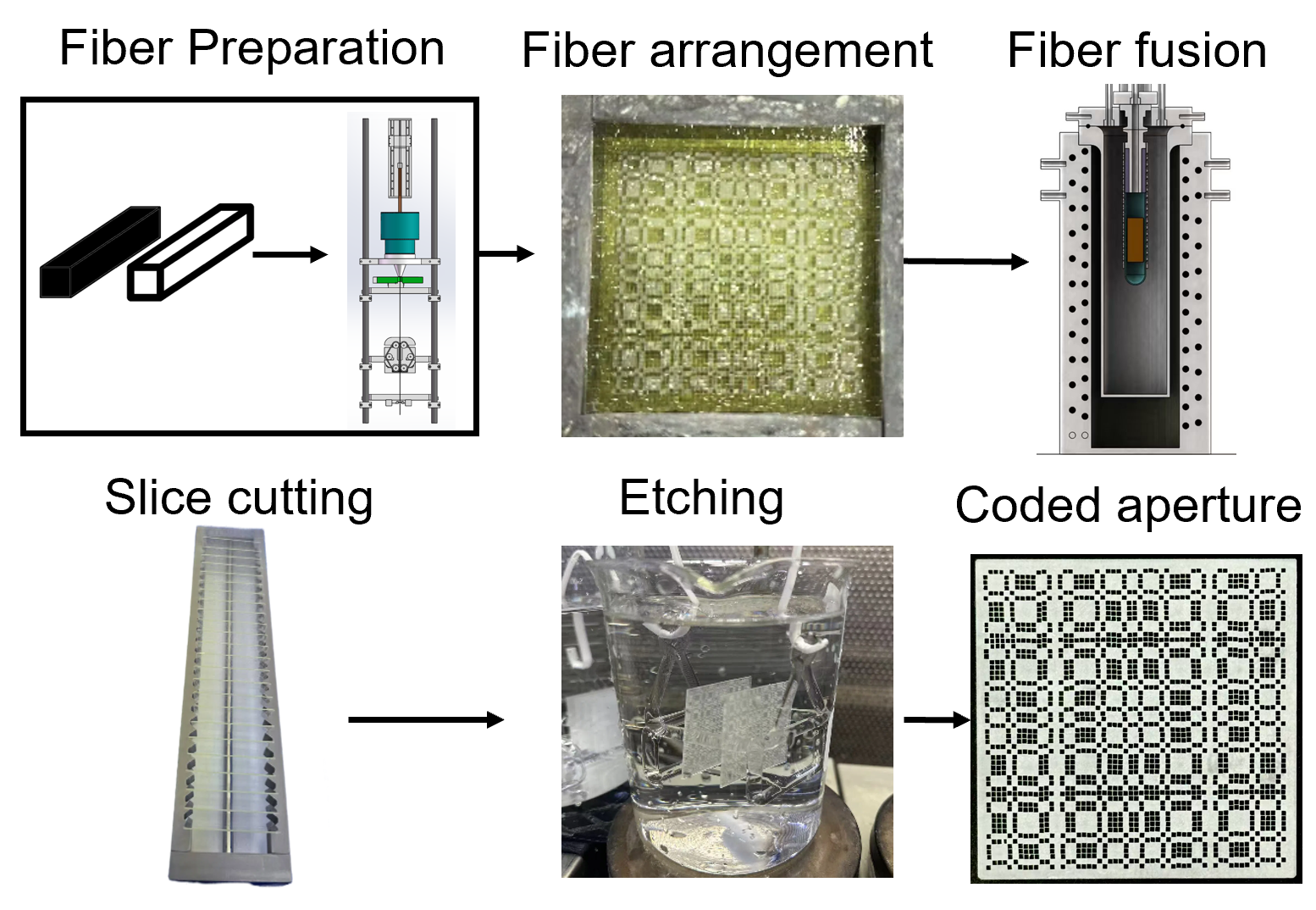}
    \caption{Manufacturing process \textcolor{black}{glass-based coded mask}.}
    \label{fig:Manufacture}
\end{figure}

  The coded mask is made of high‑lead, low‑background glass and contains 3,249 glass fibers. The manufacturing process of this coded mask is similar to that of the microchannel plate \cite{LIMCP}. As shown in Fig. \ref{fig:Manufacture}, the manufacturing process can be divided into several stages. The first step is the preparation of glass fiber. The molten glass is drawn into long, thin fibers. Glass fiber is divided into corrosion-resistant types used in the opaque area of the mask and non corrosion-resistant types used in the open area of the mask. The glass fibers are arranged in a MURA pattern and bundled tightly together. In the process of high temperature fusion, the bundled fibers are heated until the glass fuses together. The fused glass bundles are cut into slices and chemically etched to remove a portion of the material. The thickness of the glass-based coded mask is 0.9 mm, and the total weight is only 2.7 g. The opening area unit is a corroded opening of the glass fiber. The remaining material creates a self-supporting grid structure with a typical width of around 61 $\mu$m.

%%%%%%%%%%%%%%%%%%%%%%%%%%

%%%%%%%%%%%%%%%%%%%%%%%%%%
%%\begin{figure}
%%  \centering
%%  \adjustbox{valign=c}{\includegraphics[width=3.5cm]{FinalMask.png}}
%%  %\hspace{0.41cm}
%%   \adjustbox{valign=c}{\includegraphics[width=3.5cm]{SEMMASK.jpg}}
%%  \caption{Left: Photograph of the glass-based coded mask. Right: SEM micrograph of the coded mask. Typical mask feature is approximately 0.714 mm on a side and is spaced by a supporting grid that is 0.061 mm thick.}\label{fig:MASKPhoto}
%%  \vspace{\fill} 
%%\end{figure}

%\begin{figure}
%  \centering
%    \includegraphics[width=0.6\colwidthdc]{SEMMASK.jpg}
%    \caption{SEM micrograph of the coded mask. Typical mask feature is approximately 0.714 mm on a side and is spaced by a supporting grid that is 0.061 mm thick.}
%    \label{fig:MASKPhoto}
%\end{figure}

\section{X-ray Beam Test}

\par \textcolor{black}{The beam test results provide certain constraints for the mass model and energy response in subsequent Monte Carlo simulations, leading to more reliable simulation results. The purpose of the X-ray beam test is to measure the energy angular response at on-axis incidence of the detector module, thereby verifying the basic performance of the detector under actual manufacturing process conditions. }

\subsection{Experimental Setup}
\par
\begin{figure}
  \centering
    \includegraphics[width=0.9\colwidthdc]{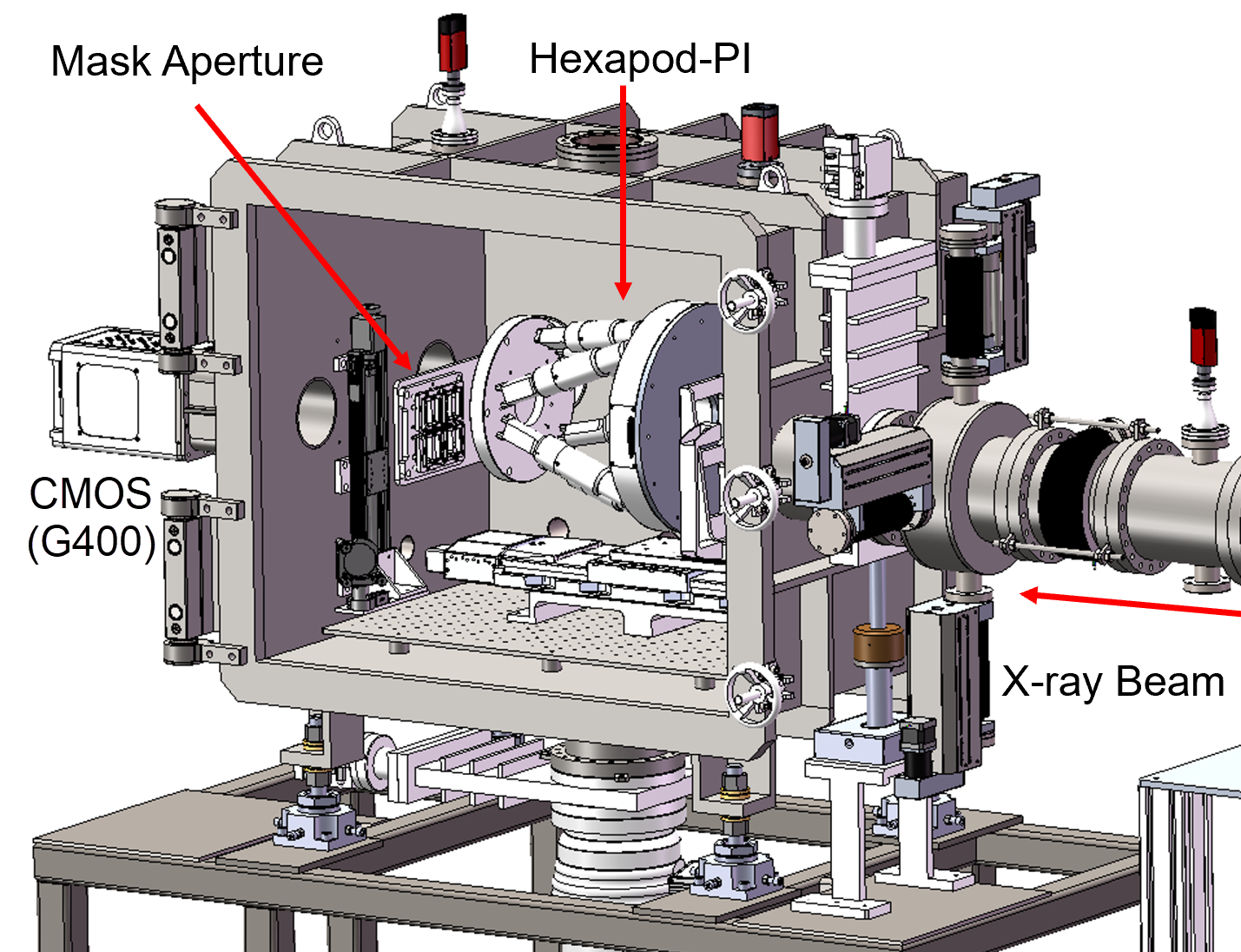}
    \caption{Experimental setup of the X-ray beam. \textcolor{black}{By measuring vertically incident X-rays, encoded images can be obtained to verify the basic functions of encoded imaging and provide constraints for subsequent performance simulations.}}
    \label{fig:BeamSetup}
\end{figure}

 The beam experiment is conducted at North Night Vision Technology Co., Ltd, and the beam facility (Fig. \ref{fig:BeamSetup}) employs an X-ray generator to produce a continuous spectrum of X-ray radiation. Irradiating the Ti target with X-rays generates characteristic emission lines at 4.51 keV and 4.93 keV. The point source (KYW500, KEYWAY ELECTRON, China) was a microfocus Ti target with a spot diameter of about 35 $\mu$m. X-rays travel through a collimator that is 14.79 meters long, where they are transformed into nearly parallel radiation. The cooling temperature of the vacuum tube for the experiment was set at -20 $^{\circ}$C. The code plate was installed on the PI six axis displacement table (H-850, six axis, Germany) at the front end for position adjustment . At the other end of the vacuum tube, a CMOS camera (G400) \cite{G400camera} measures the encoded pattern of the coded mask, with a spacing of 365 mm. This section shows the decoded image-plane intensity distribution of an on-axis X-ray source, verifying the feasibility of applying a glass-based coded mask. 
%%%%%%%%%%%%%%%%%%%%%%%%%%
%%%%%%%%%%%%%%%%%%%%%%%%%%

\subsection{Measured Energy Spectra}

\begin{figure}
  \centering
    \includegraphics[width=\colwidthdc]{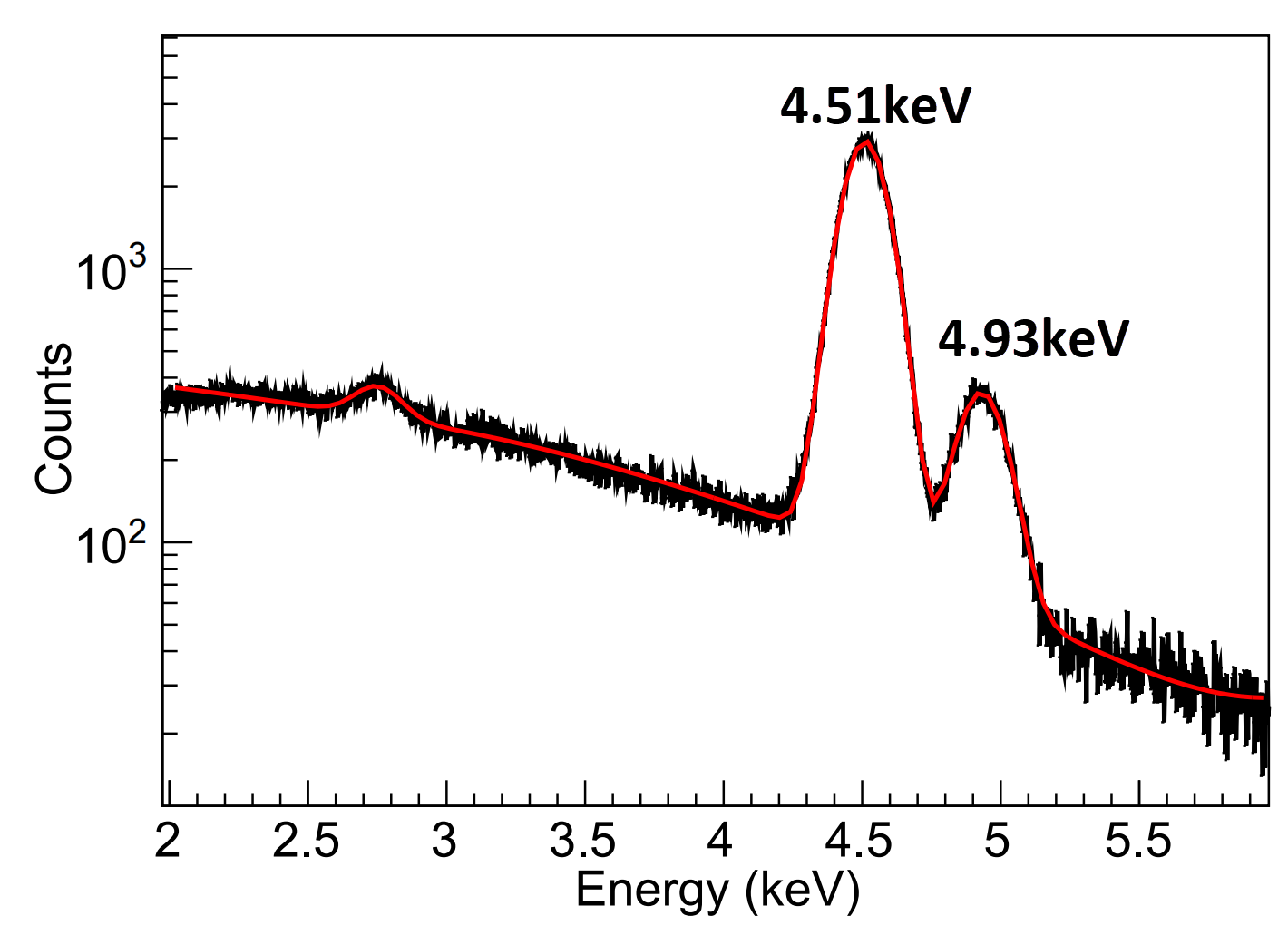}
    \caption{Spectrum of the Ti line at 4.51 keV and 4.93 keV measured by CMOS. The accumulation time is 120 seconds.}
    \label{fig:CMOSSpectrum}
\end{figure}

\par \textcolor{black}{The characteristic emission lines of Ti are employed to calibrate the ADC channel to energy. The resulting calibration function is subsequently applied to the entire raw spectrum, producing an energy spectrum. As illustrated in Fig. \ref{fig:CMOSSpectrum}, the calibrated spectrum is presented with energy on the horizontal axis and counts on the vertical axis, such that both the characteristic line features and the underlying continuum are clearly visible. The continuum background is represented by a third-order polynomial, whereas the characteristic peaks are concurrently modeled using a superposition of multiple Gaussian functions. }

\par The Gaussian standard deviation is transformed into the corresponding full width at half maximum (FWHM), from which the energy resolution is subsequently determined. The 4.51 keV and 4.93 keV lines produced by the incident X-ray beam are clearly separated from the background. Under an operating temperature of -20$^{\circ}$C, the energy resolutions (FWHM) are measured to be 199 eV at 4.51 keV and 247 eV at 4.93 keV. In addition, the spectrum exhibits an escape peak at 2.81 keV associated with the 4.51 keV line, whereas a second expected escape feature at 3.23 keV cannot be resolved.
%%%%%%%%%%%%4.93 keV 峰因计数较少，是能量分辨率误差的主要来源；4.51 keV 峰统计更好，误差贡献小得多。4.51 keV 峰（±3σ）面积：576.366   4.93 keV 峰（±3σ）面积：63.1421
%
%We constructed a mass model of the code board using geant4 and simulated the code board occlusion rate based on Monte Carlo method. The simulated data in Fig. \ref{fig:SimulateMaskRate} show that below 30 keV, glass blocks most of the X-rays, 

%%%%%%%%%%%%%%%%%%%%%%%%%%
%%%%%%%%%%%%%%%%%%%%%%%%%%
%\begin{figure}
%  \centering
%\includegraphics[width=0.7\colwidthdc]{MaskRate.png}
%    \caption{Measured mask Pattern on CMOS.}
%    \label{fig:SimulateMaskRate}
%\end{figure}
%%%%%%%%%%%%%%%%%%%%%%%%%%
%%%%%%%%%%%%%%%%%%%%%%%%%%

\begin{figure}
  \centering
    \includegraphics[width=0.8\colwidthdc]{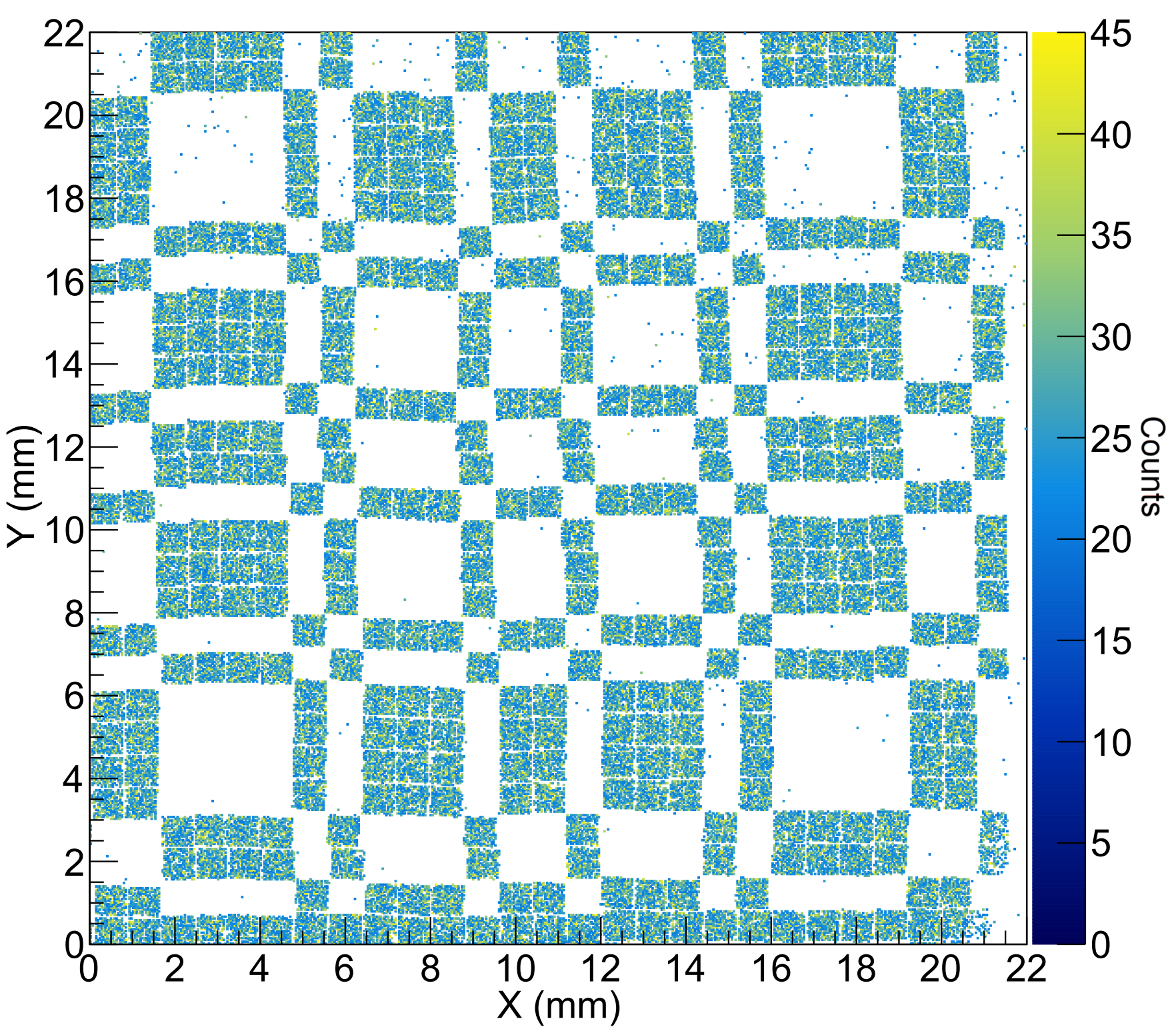}
    \caption{Mask pattern measured on CMOS. \textcolor{black}{After passing through the 14.79 m collimator, the beam generated by the X-ray tube forms an approximately parallel vertical incident beam area. A part of the spot is blocked by the coded mask, and the remaining part is projected onto the CMOS .}}
    \label{fig:MeasuredMASKPattern}
\end{figure}
\subsection{Coded Mask Decoding}
\par \textcolor{black}{The vertically incident photons pass through the code mask and leave a shadow pattern with an offset on the detector, as shown in Fig. \ref{fig:MeasuredMASKPattern}.} The measured count D of the position sensitive detector (G400) is shown in Eq.\ref{eq:eq1_label}, where S is the distribution function of the X-ray source, M is the coded mask pattern, and $\otimes$ is the correlation operator. B is the background count.
\par
\begin{figure}
  \centering
    \includegraphics[width=0.9\colwidthdc]{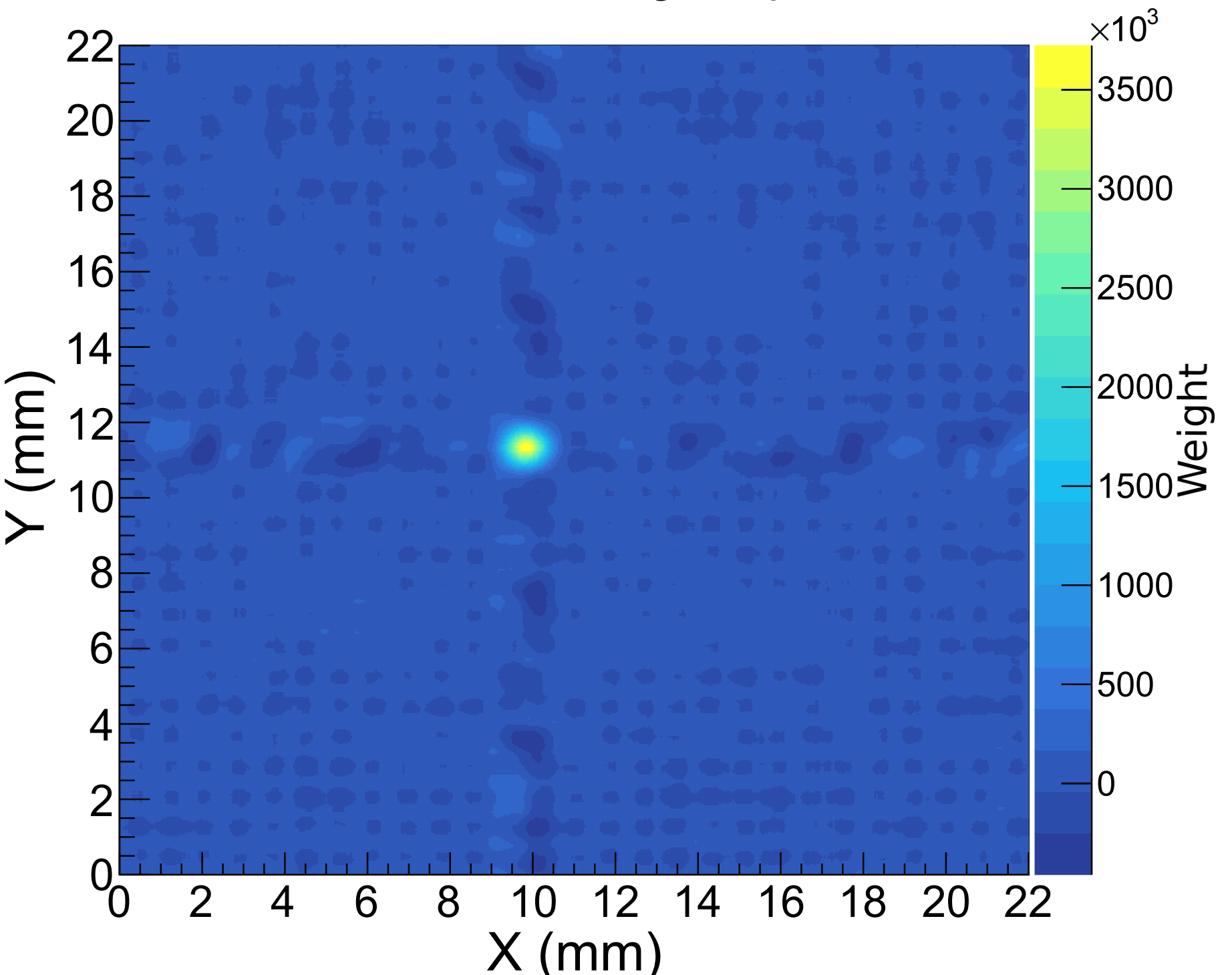}
    \caption{Decoded detector plane intensity distribution. The size of plane is 22 $\times$ 22 mm. The bright spot shows the relative position of the center of CMOS to coded mask.}
    \label{fig:BeamTestPlaneDecoding}
\end{figure}

 In order to reconstruct S from D, we use the balanced correlation of the recorded data with a suitable decoding function (Eq. \ref{eq:eq2_label}). After decoding S$_{ij}$, the detector plane intensity distribution is detected, where i and j correspond to the x and y coordinates of S, respectively. D$_{kl}$ represents the intensity distribution of the detected image, where k and l correspond to the x and y coordinates of the detected image D, respectively. The decoding functions G$^+$ and G$^-$ are related to M, as shown in Eq. \ref{eq:scaled_eq}. Because the fabricated coded masks differ slightly from their original designs, the measured mask pattern is used for the decoding process. Fig. \ref{fig:BeamTestPlaneDecoding} shows the decoded detector plane intensity distribution. The bright spot in the decoded detector plane indicates the shadow offset of the source relative to the telescope boresight.
 
\begin{equation}
    D(x,y) = S(x,y) \otimes M(x,y) + B(x,y)
    \label{eq:eq1_label}
\end{equation}

\begin{equation}
\scalebox{0.8}{$\displaystyle S_{ij} = \frac{\sum_{kl} G^+_{i+k,j+l} D_{kl}}{\sum_{kl} G^+_{i+k,j+l} } - \frac{\sum_{kl} G^-_{i+k,j+l} D_{kl}}{\sum_{kl} G^-_{i+k,j+l} }$}
\label{eq:eq2_label}
\end{equation}

\begin{equation}
\scalebox{0.8}{$
G^+ = 
\begin{cases} 
    1 & \text{if } M = 1 \\
    0 & \text{if } M = 0 
\end{cases}
\quad
G^- = 
\begin{cases} 
    0 & \text{if } M = 1 \\
    1 & \text{if } M = 0 
\end{cases}
$}
\label{eq:scaled_eq}
\end{equation}

%%%%%%%%%%%%%%%%%%%%%%%%%%%
%\begin{figure}
%  \centering
%    \includegraphics[width=0.8\colwidthdc]{XZDecode.png}
%    \caption{Typical X–Z profile plot of decoding results.}
%    \label{fig:XZProfile}
%\end{figure}
\subsection{Detector Plane to Sky-localization Map}
\par Beam source localization needs to be performed in the sky-localization map through geometric inversion. The detector-to-sky geometry follows the gnomonic (TAN) projection, with a spherical rotation from the camera frame onto equatorial coordinates \citep{Calabretta2002}. The coded mask to detector distance L is 365 mm. The conversion from the detector plane to angular coordinates $(\theta,\phi)$ is performed first.  For each pixel, its physical position $(x,y)$ is computed with respect to the image center, and r = $\sqrt{x^2+y^2}$. The telescope-direction angles are
\begin{equation}
\theta = \arctan\!\left(\frac{r}{L}\right), \qquad
\phi = \operatorname{atan2}(y,x)
\end{equation}

\par \textcolor{black}{To derive sky-localization map, it is additionally necessary to know the absolute pointing of the boresight and the orientation of the camera axes on the celestial sphere.}  The corresponding unit vector in the telescope
frame (with $z$ along the boresight, $x$ East and $y$ North) is
\begin{equation}
\mathbf{n}_{\mathrm{telescope}} =
\begin{bmatrix}
\sin\theta\cos\phi \\
\sin\theta\sin\phi \\
\cos\theta
\end{bmatrix}
\end{equation}

\par
\textcolor{black}{%
The telescope-to-equatorial transformation is the spherical Euler rotation of native coordinates about the boresight \citep{Calabretta2002}.
The equatorial basis aligned with the pointing is
{\setlength{\abovedisplayskip}{4pt}
\setlength{\belowdisplayskip}{4pt}
\setlength{\jot}{2pt}
\begin{align}
\mathbf{n}_{0}
&=
(\cos\delta_{0}\cos\alpha_{0},\;
\cos\delta_{0}\sin\alpha_{0},\;
\sin\delta_{0})
\\
\hat{\mathbf{e}}_{\mathrm{E}}
&=
(-\sin\alpha_{0},\;\cos\alpha_{0},\;0)
\\
\hat{\mathbf{e}}_{\mathrm{N}}
&=
(-\sin\delta_{0}\cos\alpha_{0},\;
-\sin\delta_{0}\sin\alpha_{0},\;
\cos\delta_{0})
\\
\mathbf{n}_{\mathrm{eq}}
&=
(n_{x},n_{y},n_{z})
=
n_{\mathrm{telescope},x}\,\hat{\mathbf{e}}_{\mathrm{E}}
+
n_{\mathrm{telescope},y}\,\hat{\mathbf{e}}_{\mathrm{N}}
+
n_{\mathrm{telescope},z}\,\mathbf{n}_{0}
\end{align}}
where $\alpha_{0}=\mathrm{RA}_{0}$ and $\delta_{0}=\mathrm{Dec}_{0}$.
$\mathrm{RA}_{0}$ and $\mathrm{DEC}_{0}$ are the telescope boresight pointing in equatorial coordinates---the sky position at which the telescope is aimed during the observation.
Assuming the telescope boresight is aligned with an observation source of $\mathrm{RA}_{0}=0$ and $\mathrm{DEC}_{0}=0$.
Then}

\begin{equation}
\mathrm{RA} = \operatorname{atan2}(n_y,n_x), \qquad
\mathrm{DEC} = \operatorname{atan2}\!\left(n_z,\sqrt{n_x^2+n_y^2}\right)
\end{equation}
\par Each pixel weight is accumulated into a two-dimensional RA–DEC histogram. The global maximum bin of the resulting sky-localization map is located at $(\mathrm{RA}_{\mathrm{peak}},\mathrm{DEC}_{\mathrm{peak}})$. \textcolor{black}{For visualization purposes, and to facilitate comparison with the true source position, we re-center the coordinate system on this maximum by plotting $(\mathrm{RA}',\mathrm{DEC}')$ in shifted coordinates such that the peak is placed at the origin in Fig.~\ref{fig:BeamTestLocalization}. This way of representing the data makes the size and orientation of the error ellipse relative to the source position easier to see.} We define the shifted coordinates
\begin{equation}
\mathrm{RA}' = \mathrm{RA} - \mathrm{RA}_{\mathrm{peak}}, \qquad
\mathrm{DEC}' = \mathrm{DEC} - \mathrm{DEC}_{\mathrm{peak}}
\end{equation}
so that the peak is at $(0,0)$.

\subsection{Gaussian Fit and Contours.}
\par \textcolor{black}{After constructing the sky-localization map, a two-dimensional Gaussian model is fitted to these discrete data points. The Gaussian centroid yields the reconstructed source coordinates in right ascension and declination, whereas the directional standard deviations and their correlation coefficient characterize the extent and orientation of the localization peak. The fit is performed via nonlinear least-squares optimization by minimizing the residuals between the Gaussian model and the observed weights evaluated at the corresponding bin centers\cite{Gros2003}}.

\par
\begin{figure}
  \centering
    \includegraphics[width=\colwidthdc]{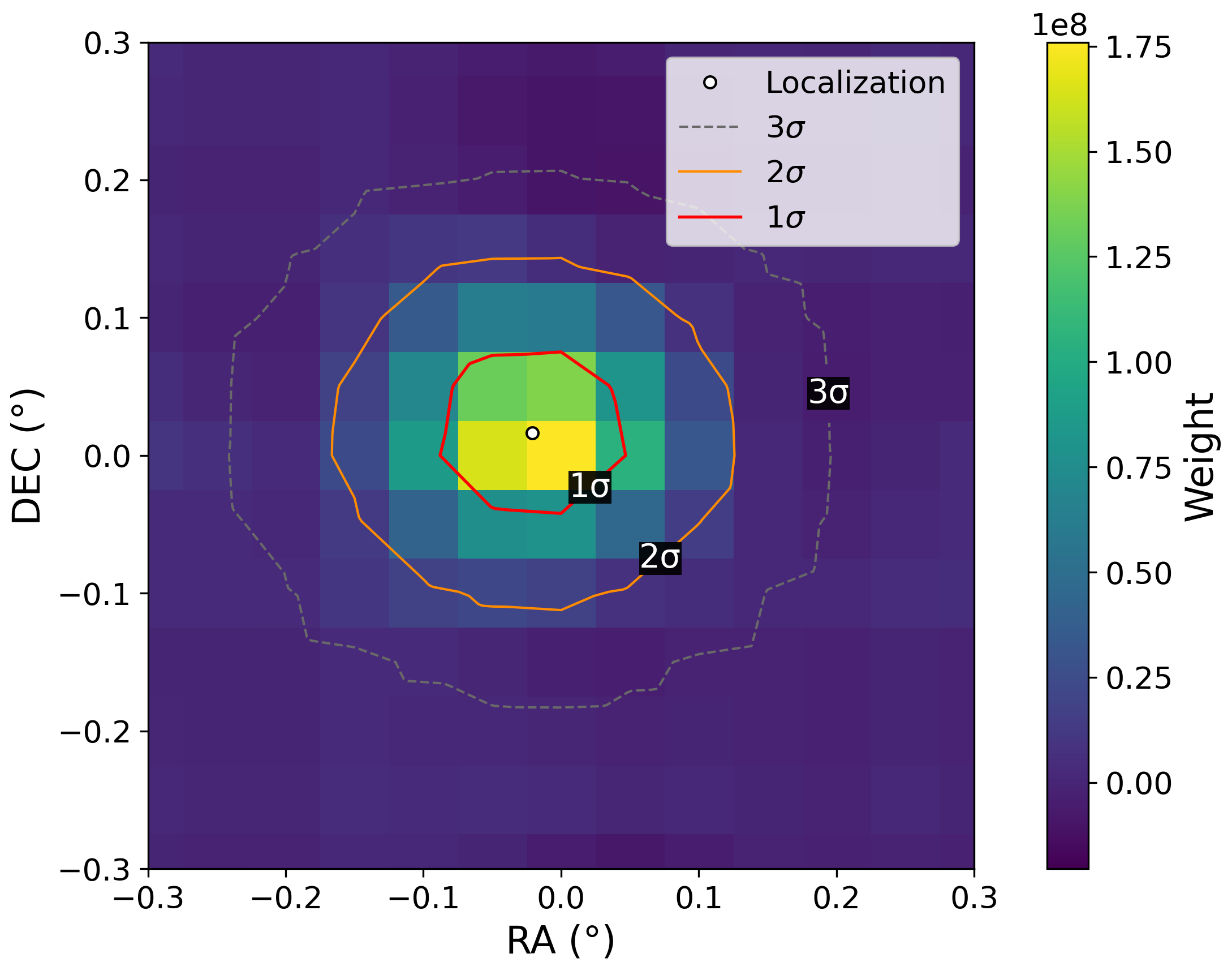}
    \caption{The localization of beam test data, the white \textcolor{black}{circle} is the localization center. The three circles from inside to outside are the 1-$\sigma$, 2-$\sigma$, 3-$\sigma$ confidence intervals, respectively.}
    \label{fig:BeamTestLocalization}
\end{figure}

 \textcolor{black}{The maximum-likelihood centroid is adopted as the source position. From the covariance matrix, we derive the semi-major and semi-minor axes of the 1$\sigma$ confidence ellipse, as well as its orientation. The projections of this ellipse onto the right ascension and declination axes yield one-dimensional estimates of the localization uncertainty. Contours are plotted which correspond to $1\sigma$, $2\sigma$, and $3\sigma$ levels. Fig. \ref{fig:BeamTestLocalization} shows the localization of beam test data. The peak-centered localization result is $\mathrm{RA}'=-0.021'$ and $\mathrm{DEC}'=0.016'$, indicating a slight offset with respect to the peak center. The fitted Gaussian 
$\sigma$ are
$\sigma_{\mathrm{maj}}=4.25'$ and $\sigma_{\mathrm{min}}=3.75'$, corresponding to an angular resolution of
$\mathrm{FWHM}_{\mathrm{maj}}=10.0'$ and $\mathrm{FWHM}_{\mathrm{min}}=8.83'$. The equivalent width is defined from the geometric-mean scale
$\sigma_{\mathrm{eq}}=\sqrt{\sigma_{\mathrm{maj}}\,\sigma_{\mathrm{min}}}$, so that
$\mathrm{FWHM}_{\mathrm{eq}}=2.355\sigma_{\mathrm{eq}}=9.40'$.}

\par \textcolor{black}{
For a coded mask imager, the angular resolution is approximated by
\begin{equation}
\Delta\theta_{\mathrm{ar}}
=
\sqrt{\left(\frac{m}{d}\right)^{2}+\left(\frac{p}{d}\right)^{2}}
\end{equation}
where $m$ and $p$ are the mask-element and detector-pixel sizes, and
$d$ is the mask-detector separation \citep{Skinner2008,Caroli1987}.
With $d=160\,\mathrm{mm}$ , $m=0.71\,\mathrm{mm}$ and $p=0.204\,\mathrm{mm}$ after
$10\times10$ binning, we obtain an angular resolution of $15'$.}

\par \textcolor{black}{The fitted FWHM (9.4$'$) and the geometric angular resolution
$\Delta\theta_{\mathrm{ar}}$ agree at the expected level $15'$.
The estimate
$\Delta\theta_{\mathrm{ar}}$
is only an approximate resolution scale, not a predicted Gaussian
width, and $m$ itself depends on how the mask-hole size is defined.
Decoding produces a smooth system point-spread function rather than
a hard-edged shadow, while binning, background, and the fit window
further shift the apparent peak. }

\subsection{Imaging SNR and Localization Uncertainty.}
\par \textcolor{black}{Following the conventional coded-mask formalism, in which the statistical uncertainty on the peak centroid is assumed to scale as the peak width divided by the imaging signal-to-noise ratio (SNR)\citep{Skinner2008,Braga2019}, we adopt
\begin{equation}
\Delta\theta_{\mathrm{loc}}
=
\frac{\mathrm{FWHM}}{\mathrm{SNR}}
\end{equation}
and report $\Delta\theta_{\mathrm{loc}}$ for the major, minor, and equivalent axes. Note that absolute pointing uncertainties and calibration-related systematic errors are explicitly excluded from this term.}

\par The imaging SNR is defined on the sky map as

\begin{equation}
\mathrm{SNR}
=
\frac{P-\mu_{\mathrm{bg}}}{\sigma_{\mathrm{bg}}}
\end{equation}

\textcolor{black}{where $P$ is the fitted peak height and
$\mu_{\mathrm{bg}}$ is the mean level in a far-field annulus that excludes
the source core. $\sigma_{\mathrm{bg}}$ is the  background noise rms. For $\mathrm{FWHM}=9.4'$ and the computed result $\mathrm{SNR}=55.12$,
this yields $\Delta\theta_{\mathrm{loc}}=0.17'$.}
%%%%%%%%%%%%%%%%%%%%%%
\section{\textcolor{black}{In-flight} Performance Estimation}
\par \textcolor{black}{The performance of the LSXI is highly dependent on its orbit, as it can be installed on nano-satellites in LEO or DRO orbits. For example, we conducted a Monte Carlo simulation using the Geant4 toolkit \cite{Geant4Agostinelli2003} and developed a complete mass model of the 6U CubeSat in a distant retrograde orbit (DRO), incorporating the LSXI.} The configuration of the LSXI in this mass model is identical to that of the structural model shown in Fig.~\ref{fig:SatelliteStructure}. The measured coded mask pattern and energy response in the beam test are imported into the mass model. 
\subsection{Instrument Response}
\par
\begin{figure}
  \centering
    \includegraphics[width=\colwidthdc]{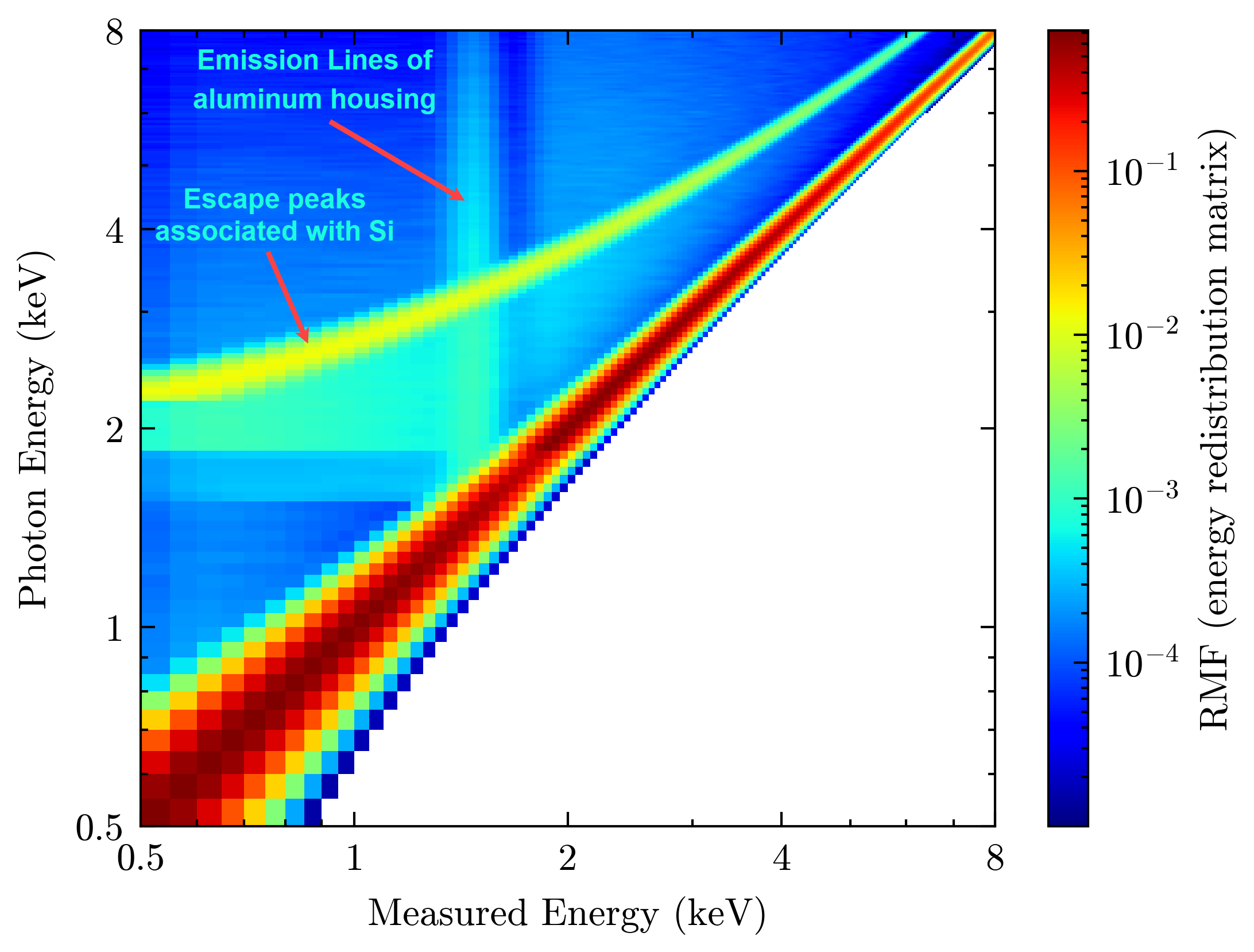}
    \caption{Energy response matrix for on-axis incidence.}
    \label{fig:RMF}
\end{figure}

 The energy response matrix is defined as the conversion between the measured spectrum and the incident spectrum. As shown in Fig. \ref{fig:RMF}, the simulated energy response matrix is plotted with an energy interval of 0.05 keV. The horizontal axis represents the energy deposited in the active region, while the vertical axis represents the incident photon energy. The color bar indicates the detection efficiency on a logarithmic scale. The diagonal area of the matrix shows the probability of full energy deposition, which is lower than the incident energy. There is a distinct line around the measured energy of 1.49 keV (K$_{\alpha}$) and 1.55 keV (K$_{\beta}$) X-rays generated when X-rays hit the aluminum shell of the detector. Another important feature that can be clearly seen is the escape peaks associated with the K$_{\alpha}$ and K$_{\beta}$ emission lines of Si (1.74 keV and 1.83 keV, respectively), which is the detection material for CMOS. 

%%%%%%%%%%%%%%%%%%%%%%%%%%

%%%%%%%%%%%%%%%%%%%%%%%%%%%%%%

\par
\begin{figure}
  \centering
    \includegraphics[width=0.9\colwidthdc]{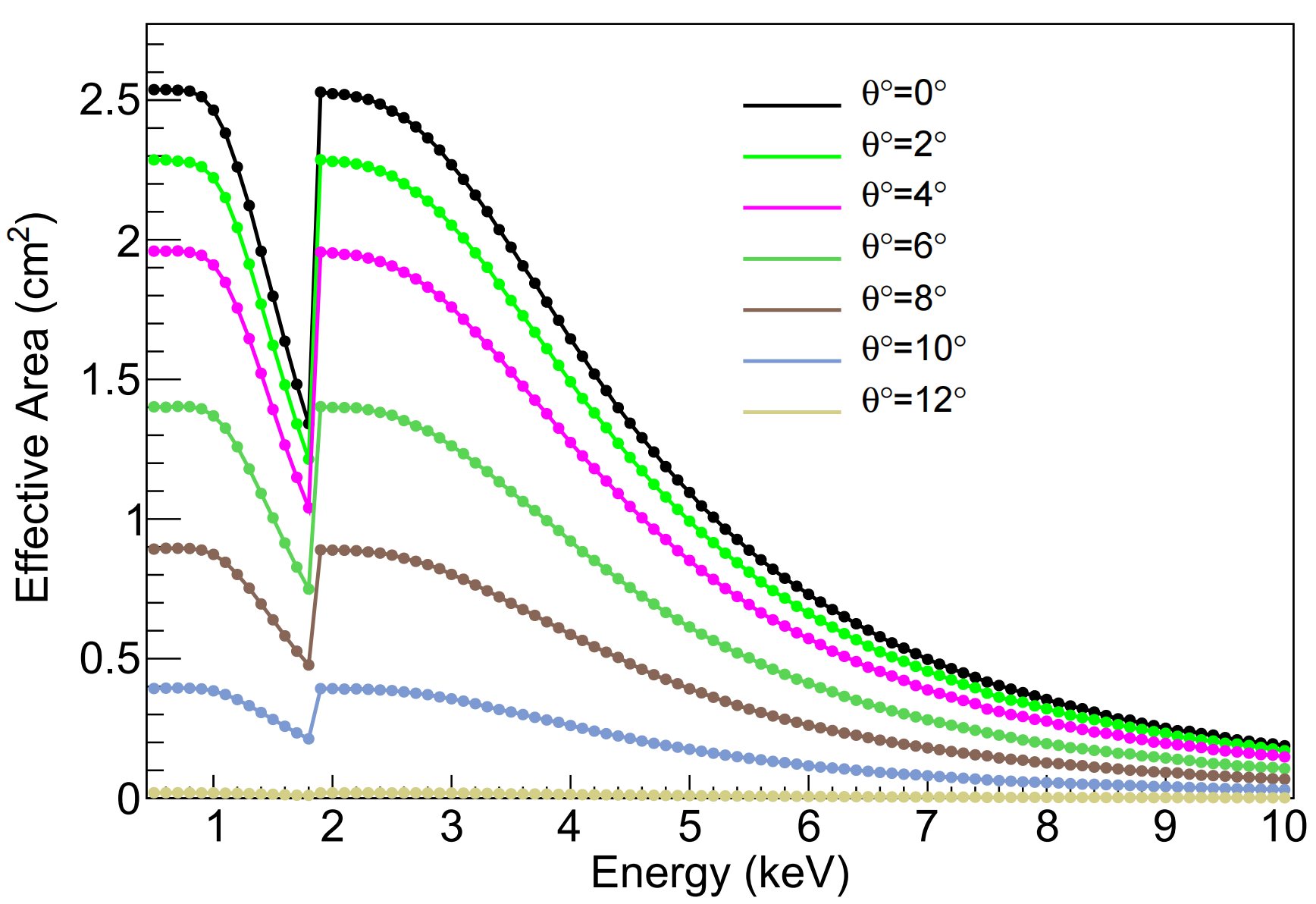}
    \caption{Effective area \textcolor{black}{of LSXI detector} as a function of energy.}
    \label{fig:SensitiveArea}
\end{figure}

 Since the photons from the X-ray sources in orbit come from all directions onto the detector, there is an effective area reduction from the actual geometric area of the detector. This results from the intrinsic detection efficiency of the CMOS and the blocking caused by the detector shielding. The total effective area as a function of energy and incidence angle is shown in Fig. \ref{fig:SensitiveArea}. Because of the absorption edge of Si, the effective area decreases markedly in the vicinity of the K$_\alpha$ emission line absorption edge. As the energy becomes higher, high-energy photons are increasingly likely to traverse the sensor without undergoing any interaction. In this way, the sensitive area shrinks. Assuming that the effective area at on-axis incidence is greater than \textcolor{black}{0.35 cm$^2$}, the detection energy range of LSXI is estimated to be within 8 keV. When the incidence angle $\theta$ is greater than the FOV of \textcolor{black}{11.65$^{\circ}$}, the detector becomes obstructed by the satellite structure.
%%根据有效面积计算灵敏度
%%%有效面积错了，需要重新改！！！！
%%%%%%%%%%%%%%%%%%%%%%%%%%
%增加能量响应矩阵的一些说明，加能量响应矩阵的图
%利用最亮伽马暴09A的样本和本底抽样的数据，做CMOS在轨图像的模拟，并对正入射的解码图像和角分辨进行分析。

\subsection{In-flight Background}
% Requires: \usepackage{amsmath}
\par The sensitivity is determined from simulations of the effective area combined with a model of the expected background in the DRO orbit. Although the in-orbit background can be partially shielded by the satellite structure and the detector’s 1 mm aluminum shielding, it remains the dominant factor determining the performance of LSXI. \textcolor{black}{Compared with cosmic rays (mainly protons) and the cosmic X-ray background (CXB) \cite{CXBGiacconi1962}, the fluxes of electrons, heavy nuclei, and positrons are typically much lower \cite{wang_2024_GECAMDBK} and are therefore neglected in this work. In addition, activation of structural materials, particularly due to charged particles (mainly protons) in orbit, can also contribute to the background, which is called the delay background.} The SPENVIS online program \cite{spenvis} is used to generate particle fluxes based on the orbital parameters. The CREME86 model is adopted with a one-year averaging period to estimate the mean particle flux in the DRO orbit. \textcolor{black}{Solar energetic particles produced during periods of solar activity can also enhance the X-ray background in space; however, such intervals are typically excluded in scientific analyses. Therefore, the solar contribution to the LSXI background is not explicitly considered in this study}. Consequently, only the CXB and protons are included in the simulations.

\par
\begin{figure}
  \centering
    \includegraphics[width=0.9\colwidthdc]{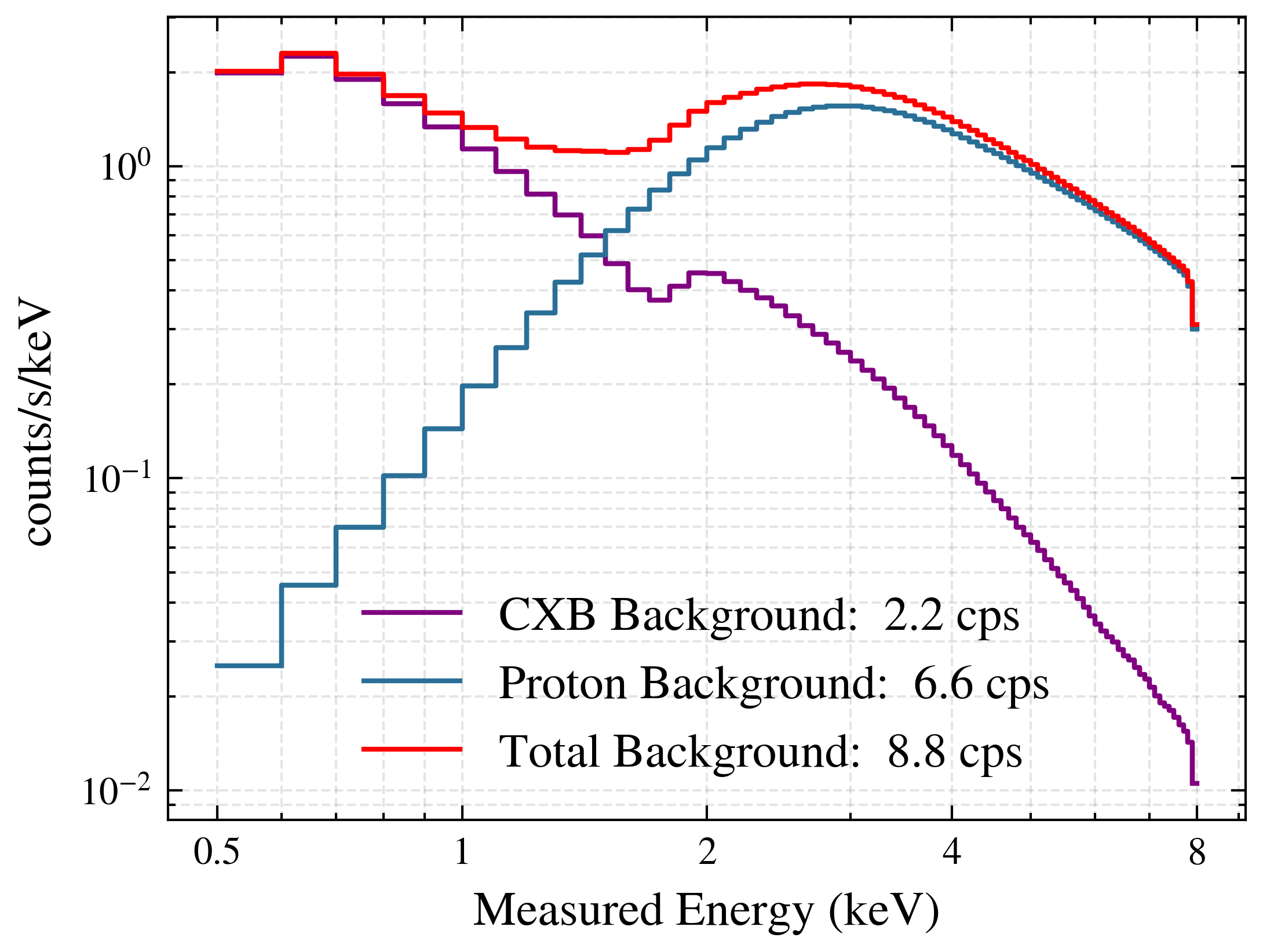}
    \caption{Background simulations based on the CXB and CREME86 models. \textcolor{black}{The detector housing is made of 1mm thick aluminum, which can shield a portion of the on-orbit background.}}
    \label{fig:BK}
\end{figure}

 The simulation of the measured background is shown in Fig. \ref{fig:BK}.  The energy distribution of the CXB is shown in Eq. \ref{eq:CXBBK}. Although the detection range of the LSXI is 0.5-8 keV, for the CXB background, as the incident energy widens, low-energy scattered photons are generated due to compton scattering in the target material and are detected by the CMOS. Therefore, the energy range of the CXB background input is selected as 0.5 keV to 1 MeV. Because the detector’s structure blocks low-energy protons, the input energy range is chosen to be 1–80,000 MeV. \textcolor{black}{The CXB background dominates in the low-energy band, and the proton background increases in the high-energy band.}

\begin{equation}
F(E) = \begin{cases}
0.54 E^{-1.4}, & E<0.02 MeV,\\[1pt]
0.0117 E^{-2.38}, & 0.02 MeV\leq E<0.1 MeV,\\[1pt]
0.014 E^{-2.3}, & E\geq0.1 MeV.
\end{cases}
\label{eq:CXBBK}
\end{equation}

%%%线换颜色，红绿蓝
%%%%%%%%%%%%%%%%%%%%%%%%%%

\subsection{Sensitivity}

\par \textcolor{black}{We employ rate trigger to quantify the detection sensitivity. Rate triggers exhibit enhanced sensitivity to short-duration bursts and brief pulses relative to image triggers, because they respond to a rapid excess in count rate rather than relying on a prolonged image integration\cite{Band2006SwiftSensitivity,Fenimore1978}. They also operate on shorter timescales, which facilitates more rapid follow-up observations. Image triggers, however, remain superior for detecting long-lasting, low-flux, or slowly rising events, and rate triggers typically still require subsequent imaging for accurate localization and rejection of spurious triggers.} 
\par Under the conditions of Poisson statistics, the statistical significance of detection is described by N$_{\sigma}$ = S/$\sqrt{N}$. S is the total source counts detected during the observation. N is the total background counts detected by the CMOS detector, which can be determined by N = B$\cdot$$A_{\mathrm{geo}}$$\cdot$$\Delta$E$\cdot$T. B is the detected background counts in counts cm$^{-2}$s$^{-1}$keV$^{-1}$. $A_{geo}$ is the geometric area of the detector, ${\Delta E}$ is the energy range under consideration (in keV), and T is the exposure time. The total counts of the source (S) are given by Eq. \ref{eq:SourceCounts}, where $F_{\min}$ is the minimum detectable source flux and $A_{\mathrm{eff}}$ is the effective area derived from the simulations shown in Fig. \ref{fig:SensitiveArea}.

\begin{figure}
  \centering
    \includegraphics[width=\colwidthdc]{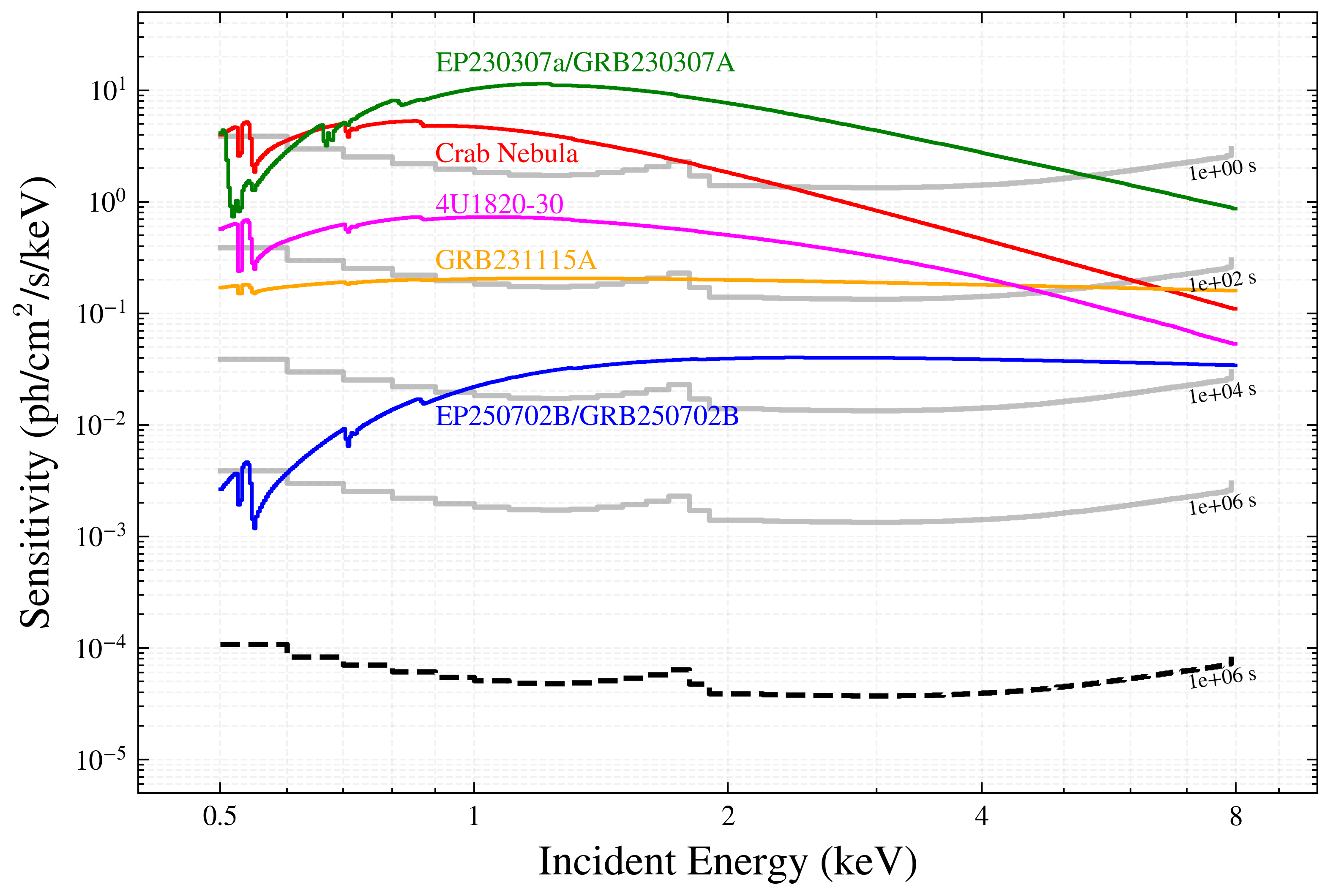}
    \caption{\textcolor{black}{Rate trigger is used to quantify the detection sensitivity. 3$\sigma$ continuum sensitivity for different exposure times for LSXI with a single module (gray lines) and an array of multi-modules with a total detection area of 5200 cm$^2$ (dotted black line)}. We simulate observations of several representative astrophysical sources, including the Crab Nebula \cite{2023A&A...671A..67D}, Gamma-ray bursts (GRBs) (GRB 230307A, GRB 250702B) \cite{2024cosp...45.1786S, 2026ApJ...997L..45Z}, fast X-ray transients (FXTs) (EP 230307a, EP250702a) \cite{2024cosp...45.1786S, 2024GCN.38081....1Z, 2025ApJ...987L..38Z}, magnetar giant flares (GRB 231115A) \cite{2024Natur.629...58M, 2024ApJ...969..127W}, tidal disruption events (TDEs) (Ep 250702a) \cite{2026ApJ...997L..45Z}, and LMXB (4U1820–30)  \cite{2025A&A...697A..83A}. Ccontinuum sensitivity is calculated assuming $\Delta E = 0.5E$\cite{2022arXiv221116916V}}
    \label{fig:Sensitivity}
\end{figure}

\begin{equation}
%\footnotesize
\begin{split}
    S = F_{\text{min}}\cdot A_{\text{eff}}   \cdot\Delta E\cdot T 
\end{split}    
    \label{eq:SourceCounts}
\end{equation}

% Requires: \usepackage{amsmath}
\begin{equation}
%\footnotesize
\begin{split}
    F_{\min} 
    & = \frac{N_{\sigma}\,}{A_{eff}  } \sqrt{\frac{B\cdot A_{geo}\,}{\Delta E\cdot T}} \, photons\, cm^{-2}s^{-1}keV^{-1}
\end{split}
    \label{eq:Sensitivity}
\end{equation}

\par Therefore, the continuum sensitivity at a given energy can be expressed by Eq. \ref{eq:Sensitivity} \cite{Braga2020}. Assuming N$_{\sigma}$ = 3, T = 1, $10^2$ , $10^4$ s, and $10^{6}$ s, we plot the on-axis continuum sensitivities in Fig.\ref{fig:Sensitivity}. For an observation time of \(T = 10^{6}\,\mathrm{s}\), the corresponding sensitivity for a single telescope module is 1.86 mCrab. For comparison, we also overplot the incident fluxes of several representative classes of astrophysical transients and variable sources, including gamma-ray bursts (GRBs) \cite{2024cosp...45.1786S, 2026ApJ...997L..45Z}, fast X-ray transients (FXTs) \cite{2024GCN.38081....1Z, 2025ApJ...987L..38Z}, magnetar giant flares (MGFs) \cite{2024Natur.629...58M, 2024ApJ...969..127W}, tidal disruption events (TDEs) \cite{2026ApJ...997L..45Z},  NS low-mass X-ray binaries  (LMXB) \cite{2025A&A...697A..83A}, with interstellar absorption modeled using \texttt{tbabs} \cite{2000ApJ...542..914W}. Although the effective area of a single module is insufficient to detect relatively faint sources, the sensitivity can be improved by combining multiple identical modules. In the background-dominated regime, the limiting flux scales approximately as 1/$\sqrt{n}$, where n is the number of modules. The dotted black curve in Fig. \ref{fig:Sensitivity} illustrates the expected sensitivity of LSXI for an equivalent total detection area of 5200 cm$^2$, corresponding to the detector area of \textit{Swift}/BAT, and is shown for illustrative comparison only. The corresponding sensitivity with an observation time of \(T = 10^{6}\,\mathrm{s}\) is 0.05 mCrab.
%说明一下，探测模块增加后，灵敏度的变化趋势，列出灵敏度变化的公式Fmin’（N）= Fmin / sqrt（N）??,增加一根100模块的灵敏度示意图。
%增加一些伽马暴亮度的点，确认下可以观测到什么源
% 背景主导下，灵敏度随 A_eff 增大而提高，随曝光时间 T 增大而按 1/sqrt(T) 改善
% 多个相同模块组合后，灵敏度提升约为 1/sqrt(n)

%\begin{equation}
%\footnotesize
%    \sigma = \frac{S}{\sqrt{N}}
%    \label{eq:significance}
%\end{equation}

% Requires: \usepackage{amsmath}
%\begin{equation}
%\footnotesize
%    N = B(E) \cdot \Delta E\cdot \Delta t
%    \label{eq:NoiseCounts}
%\end{equation}

% Requires: \usepackage{amsmath}

%%%%%%%%%%%%%%%%%%%%%%%%%%加100s 的灵敏度，同时换算一下是多少mCRAB？？？？

%\par A series of monoenergetic far-field line sources from the CXB model and the GRB~221009A spectrum model determined by GECAM-C are simulated on-axis, as described in Eq.\ref{eq:Band} and Eq.\ref{eq:CXB}.
%%%%%%%%%%%%%%%%%%%%%%%%%%%%%%%%%%%%%%%%%%%%%%%%%%%%%%%%%%
\subsection{Localization}
\par To evaluate the on-axis localization capability, we simulated the pointed observation of the Crab Nebula. The Crab Nebula, a bright non-thermal emitter spanning nearly the entire electromagnetic spectrum, has long served as a standard calibration source in high-energy astrophysics. Measurements by COMPTEL and NuSTAR show that its spectrum over 0.2 keV–30 MeV is well represented by a Band-like model, with evidence for a spectral break above $\sim$ 1 MeV. In the LSXI energy range, the spectrum is adequately described by a single power-law model (see Eq. \ref{eq:CrabSpectrum}). According to Eq.~\ref{eq:CrabSpectrum}, the Crab Nebula spectrum is generated by sampling the modeled energy distribution. The cosmic X-ray background (CXB) and the spatial proton-induced background are then incorporated into the simulation. 

% Requires: \usepackage{amsmath}
\begin{equation}
    N(E) = 8.77 E^{-2.1} \, \text{photons cm}^{-2} \, \text{s}^{-1} \, \text{keV}^{-1}
    \label{eq:CrabSpectrum}
\end{equation}

\begin{figure}
  \centering
    \includegraphics[width=0.8\colwidthdc]{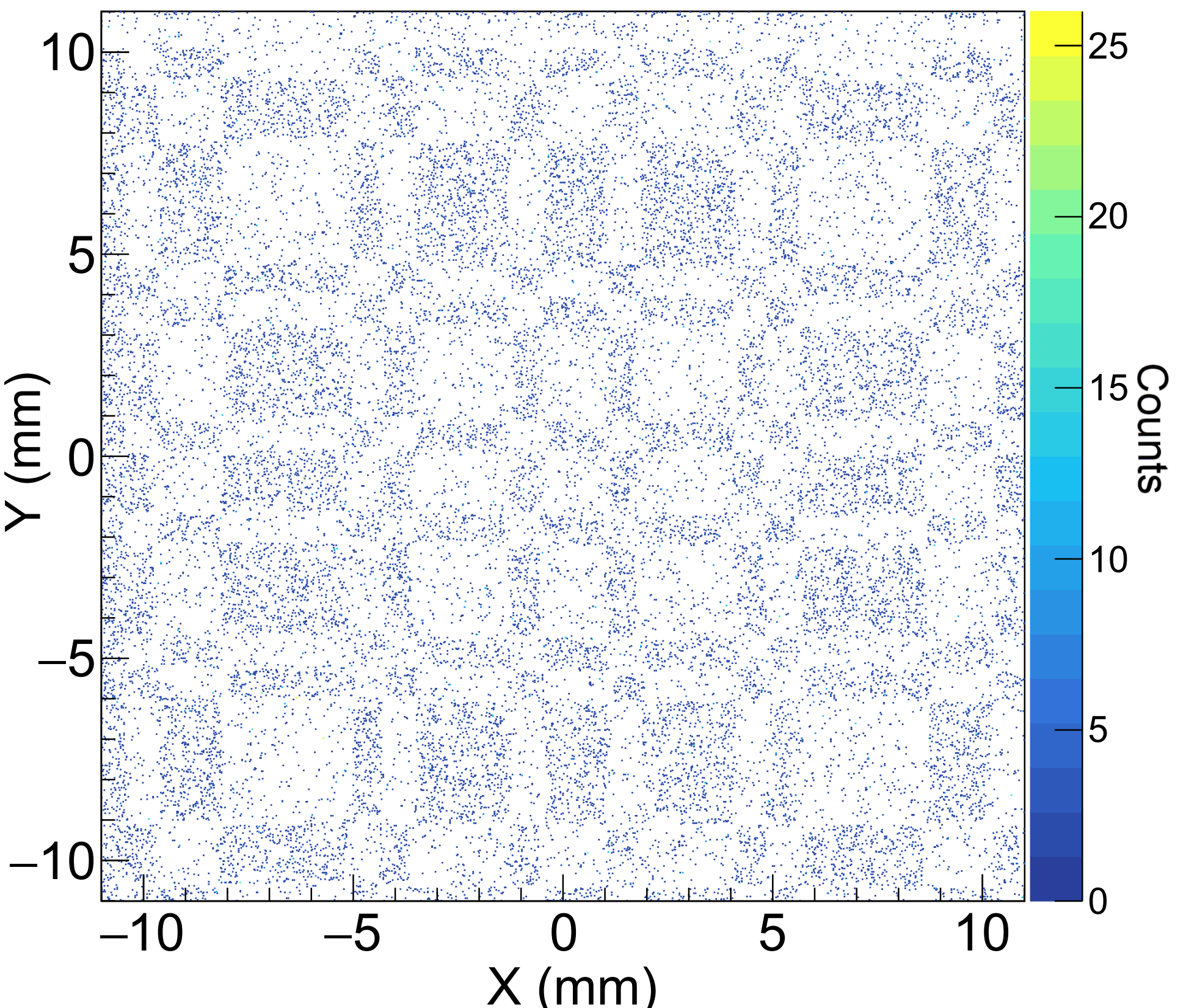}
    \caption{The simulated CMOS patten with on-axis observation of Crab Nebula. The \textcolor{black}{exposure time} is 100 s.}
    \label{fig:SimulationOfCMOSimage}
\end{figure}
%%%%%%%%%%%%%%%%%%%%%%%%%%
\begin{figure}
  \centering
    \includegraphics[width=\colwidthdc]{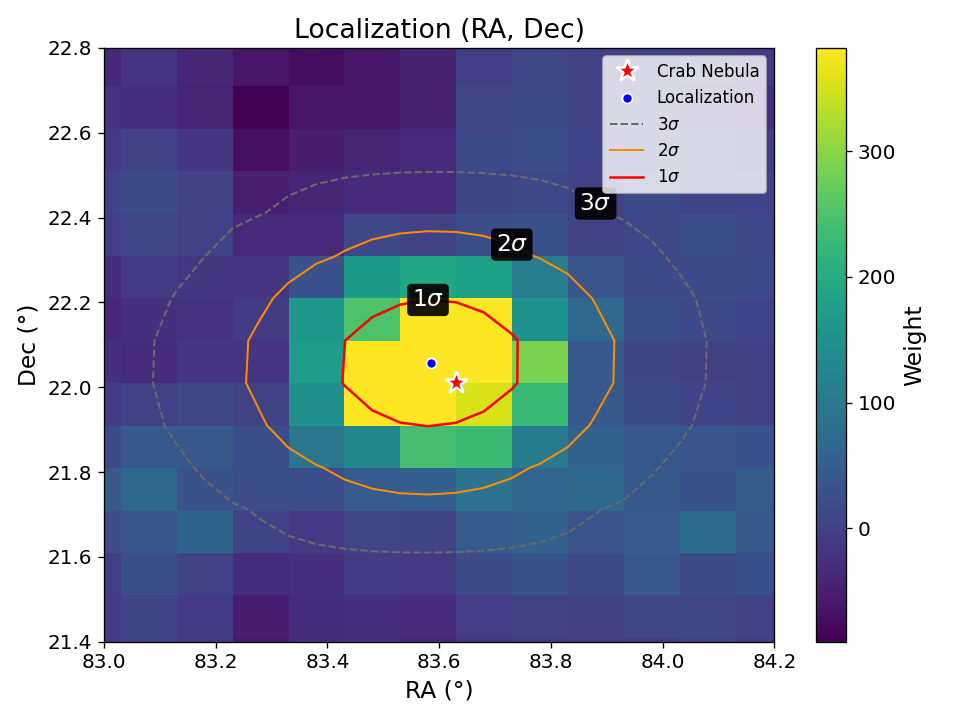}
    \caption{The simulated localization of Crab Nebula, the red and blue markers are the Crab center and localization center, respectively. The three circles from inside to outside are the 1-$\sigma$, 2-$\sigma$, 3-$\sigma$ confidence intervals, respectively.}
    \label{fig:Localization}
\end{figure}

%%%%%%%%%%%%%%%%%%%%%%%%%%
\begin{figure}
  \centering
    \includegraphics[width=\colwidthdc]{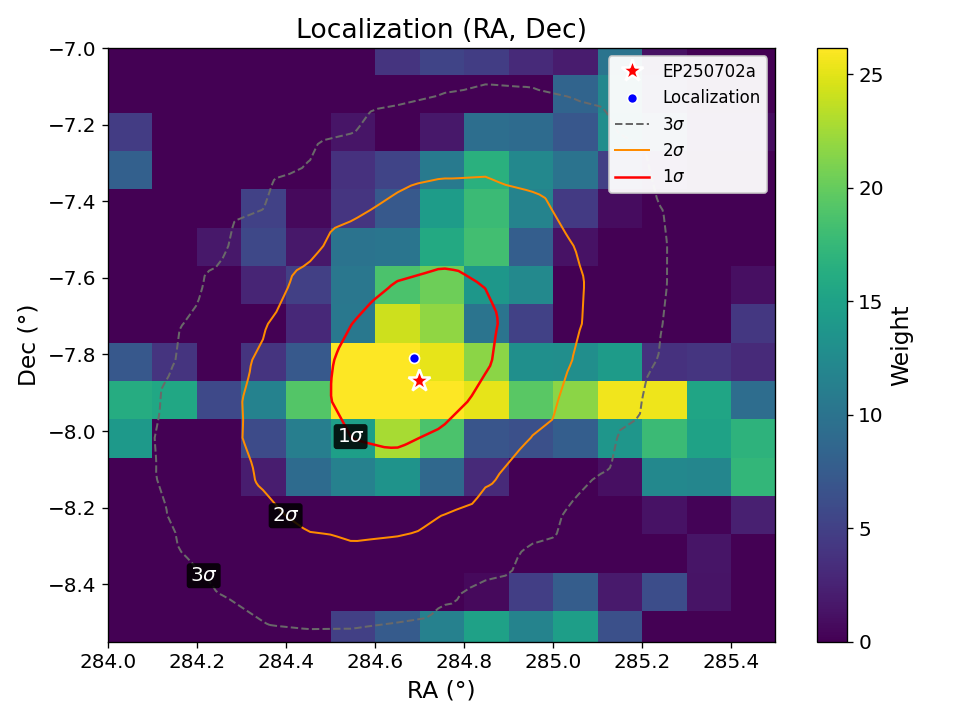}
    \caption{The simulated localization of EP250702a. The 6$^{th}$ spectral energy distributions is selected for simulation and the duration is 827 s\cite{li2026fast}}.
    \label{fig:702BLocalization}
\end{figure}

%-------- Localization (2D Gaussian fit) --------
%  Center:  RA = 83.858303 deg,  Dec = 21.776369 deg
%  sigma_RA (1-sigma projection):  0.189176 deg  =  11.3506 arcmin
%  sigma_Dec (1-sigma projection): 0.346627 deg  =  20.7976 arcmin
%  1-sigma ellipse:  sigma_maj = 0.366435 deg (21.9861 arcmin),  sigma_min = 0.147185 deg (8.8311 arcmin)
%  Overall 1-sigma (quadrature):   0.394889 deg  =  23.6934 arcmin
%  Equivalent circle (geom mean): 0.232236 deg  =  13.9342 arcmin
%\subsection{Localization Accuracy}
\par The detected total count rates of the Crab Nebula, CXB, and proton background are 15.16, 2.2, and 6.6 counts/s, respectively. The cumulative time \textcolor{black}{pointed observation} is 100 s. The measured CMOS pattern is shown in Fig. \ref{fig:SimulationOfCMOSimage} with an energy range of 0.5-8 keV. Following the procedure outlined in Section 4 and adopting a detector-to-image plane separation \(L = 16\,\mathrm{cm}\) for the 6U CubeSat, the distribution in the image plane is mapped into equatorial coordinates, namely right ascension (RA) and declination (Dec), as shown in Fig.~\ref{fig:Localization}. The fitted localization is $(83.585^\circ,22.058^\circ)$,
while the nominal true position is $(83.630^\circ,22.010^\circ)$. The corresponding angular offset is $3.81'$. The fitted angular FWHM is $22.2'$, and the SNR is 4.57. Compared with the statistical localization uncertainty $\Delta\theta_{\mathrm{loc}}=\mathrm{FWHM}/\mathrm{SNR}=4.85'$\citep{Skinner2008,Braga2019}, the offset is consistent with the $1\sigma$ statistical error.

\par \textcolor{black}{To assess the statistical localization uncertainty of LSXI for representative X-ray transient sources, we performed simulations of an on-axis observation of EP250702a \cite{li2026fast}. During the flaring episodes of EP250702a, we selected the sixth spectral energy distribution (SED6), with a duration of 827 s. The corresponding source count rate detected by LSXI is only 0.41 counts/s, which should be compared with a background rate of 8.8 counts/s. The resulting overall detection significance for EP250702a is therefore $\sim4\sigma$, demonstrating that the source is detectable on the 827 s timescale. Decoding and localizing the simulated LSXI data yields the result shown in Fig.~\ref{fig:702BLocalization}. The fitted localization is $(284.695^\circ,-7.809^\circ)$,
while the nominal true position is $(284.700^\circ,-7.869^\circ)$. The corresponding angular offset is $3.6'$. The fitted FWHM is $29.9'$ and SNR is 3.12.  Compared with the statistical localization uncertainty
$\Delta\theta_{\mathrm{loc}}=\mathrm{FWHM}/\mathrm{SNR}=9.6'$,
the residual is within the $1\sigma$ statistical error.} \textcolor{black}{Owing to the substantially lower flux of EP250702a relative to the Crab Nebula, the achieved statistical localization uncertainty is accordingly degraded.}

\section{Summary and Perspective}
\par This article presents a new glass-based coded-mask telescope module for soft X-ray imaging that satisfies the stringent mass and volume constraints of nano-satellites. A fabrication approach based on glass micro-channel plates is introduced to produce a lightweight coded mask. The practicality of the detector design is initially demonstrated by reconstructing encoded images from measurements at the X-ray beamline at North Night Vision Technology Co., Ltd. 

\par Monte Carlo simulations of the module integrated into a 6U CubeSat are conducted to assess its in-orbit detector performance, including sensitivity and source localization. 
% The results indicate the detection capability for bright astrophysical sources (such as GRBs and magnetar giant flares). 
\textcolor{black}{Our sensitivity estimates indicate that a single LSXI module is not sensitive enough to detect millisecond magnetar giant flares, owing to their extremely short durations, nor the soft X-ray emission from moderately bright GRBs on timescales of $\sim$ 10$^2$ s. Sources with even lower fluxes are correspondingly inaccessible in the single-module configuration. However, long-duration transients, such as LMXB outbursts lasting from days to tens of days, may still be detected through pointed observations with sufficiently long integrations. This suggests that the detectability of weak sources depends not only on the collecting area, but also strongly on the characteristic duration of the event and the adopted observing strategy. }
\par \textcolor{black}{Therefore, the present sensitivity mainly favors bright soft X-ray transients, while weaker short-timescale events require a larger collecting area, and weaker long-timescale events are more effectively addressed through long pointed observations. The successful coded mask fabrication, experimental validation, and performance simulations provide a strong basis for future in-orbit deployment in soft X-ray astrophysical observations and offer a new lightweight payload option.}

\par Subsequent research will focus on optimizing the fabrication process of the glass-based coded mask to minimize distortion from the designed MURA pattern and improve angular resolution. The development of a standalone detector module with a complete structure and electronic read-out system is also on the schedule. Multi-module array designs will be developed alongside advanced data analysis and decoding algorithms to improve sensitivity and localization accuracy.

\section*{Acknowledgments}
\par \emph{This work was supported by grants from the National Natural Science Foundation of China (NSFC) (grant Nos. 12173038,42474222, 12273042,12573105) and the Strategic Priority Research Program of the Chinese Academy of Sciences (grant Nos. XDA30050100, XDA30050000, XDA15360102).}

%\bibliographystyle{cas-model2-names}

\bibliography{cas-refs}

%\vskip3

\end{document}